\documentclass[%
 reprint,
 amsmath,amssymb,
 aps,
]{revtex4-2}

\usepackage{graphicx}
\usepackage{dcolumn}
\usepackage{bm}
\usepackage{hyperref}
\usepackage{amsmath,amssymb,amsfonts}
\usepackage[italicdiff]{physics}
\usepackage{cancel}
\usepackage{soul}
\usepackage[dvipsnames]{xcolor}
\usepackage{soul,xcolor}

\setstcolor{red}

\newcommand{\pr}[1]{\left(#1\right)}
\newcommand{\corch}[1]{\left[#1\right]}
\newcommand{\calF}{\mathcal{F}}
\newcommand{\eq}{\text{eq}}
\newcommand{\intern}{\text{int}}
\newcommand{\zerocurv}{\text{ZC}}
\newcommand{\BuckledStable}{\text{B+}}
\newcommand{\BuckledUnstable}{\text{B-}}
\newcommand{\eps}{\varepsilon}

\newcommand{\twodim}{\text{2d}}
\newcommand{\onedim}{\text{1d}}
\newcommand{\gs}{\text{gs}}

\begin{document}

\preprint{APS/123-QED}

\title{Buckling in a rotationally invariant spin-elastic model}

\author{Gregorio García-Valladares}
 \email{ggvalladares@us.es}
\author{Carlos A.~Plata}%
 \email{cplata1@us.es}
\author{Antonio Prados}
 \email{prados@us.es}
\affiliation{%
 Física Teórica, Universidad de Sevilla, Apartado de Correos
  1065, E-41080 Sevilla, Spain}%

\date{\today}

\begin{abstract}
Scanning tunneling microscopy experiments have revealed an spontaneous rippled-to-buckled transition in heated graphene sheets, in absence of any mechanical load. Several models relying on a simplified picture of the interaction between elastic and internal, electronic, degrees of freedom have been proposed to understand this phenomenon. Nevertheless, these models are not fully consistent with the classical theory of elasticity, since they do not preserve rotational invariance. Herein, we develop and analyse an alternative classical spin-elastic model that preserves rotational invariance while giving a qualitative account of the rippled-to-buckled transition. By integrating over the internal degrees of freedom, an effective free energy for the elastic modes is derived, which only depends on the curvature. Minimisation of this free energy gives rise to the emergence of different mechanical phases, whose thermodynamic stability is thoroughly analysed, both analytically and numerically. All phases are characterised by a spatially homogeneous curvature, which plays the role of the order parameter for the rippled-to-buckled transition, in both the one- and two-dimensional cases. In the latter, our focus is put on the honeycomb lattice, which is representative of actual graphene. 
\end{abstract}

\maketitle


\section{Introduction}\label{sec:intro}

Behaviour of low-dimensional systems is a prolific research field~\cite{cea_numerical_2020,amnuanpol_buckling_2021,chen_spontaneous_2022,hanakata_thermal_2021,hanakata_anomalous_2022,jain_compression-controlled_2021,poincloux_bending_2021,le_doussal_thermal_2021,thibado_fluctuation-induced_2020}. Two-dimensional materials exhibit intriguing properties~\cite{novoselov_electric_2004,neto_electronic_2009,samy_review_2021},  such as flexibility, high thermal and electrical conductivity, transparency, low Joule effect, to name just a few. These relevant features  can be useful for many applications, from  realisation of microsystems for healthcare~\cite{huang_graphene-based_2019,zhang_scalably_2022} up to energy harvesting~\cite{thibado_fluctuation-induced_2020,mangum_mechanisms_2021,gikunda_array_2022}. For a recent review on the effects of strains in graphene and other two dimensional materials, see Ref.~\cite{amorim_novel_2016}.

An outstanding phenomenon in the context of mechanical properties of elastic systems is the buckling transition. In structural engineering, buckling is the abrupt change of the considered structure, from a flat  to a non-zero curvature (buckled) state, under load. A prototypical example occurs when a long column, i.e. a one-dimensional system, buckles under axial compression, which is known as Euler buckling~\cite{landau_elasticity_1986,golubovic_dynamics_1998}. Nevertheless, not only is buckling an engineering subject but also an appealing physical behaviour to be better understood from a theoretical standpoint. Moreover, phenomena similar to buckling emerge in a wide range of systems, from low-dimensional systems \cite{hanakata_thermal_2021,singh_rippling_2015,plummer_buckling_2020,shankar_thermalized_2021} to cells~\cite{luo_buckling_2007} or in device development~\cite{hu_axial_2017}. 

Graphene sheets are not perfectly two-dimensional. In fact, they display transversal displacements (ripples) due to thermal fluctuations~\cite{meyer_structure_2007,fasolino_intrinsic_2007} that stabilise the membrane~\footnote{It is important to recall that the Mermin-Wagner Theorem~\cite{mermin_absence_1966} forbids the existence of purely two-dimensional stable crystals. Also, the instability of two-dimensional crystals can be understood within the context of Landau's theory of fluctuations~\cite{landau_statistical_2013}.}. Rippling has been studied employing models where its origin and behaviour are linked with the coupling between electronic and elastic degrees of freedom, i.e. the electron-phonon coupling~\cite{cea_numerical_2020,fasolino_intrinsic_2007,san-jose_electron-induced_2011,bonilla_model_2012,bonilla_ripples_2012,ruiz-garcia_stm-driven_2016,ruiz-garcia_bifurcation_2017}. Rippling has also been investigated by employing phenomenological models such as the Helfrich model~\cite{wei_bending_2013}, originally developed for membranes~\cite{helfrich_elastic_1973,lipowsky_conformation_1991}. In this approach, the free energy of the system is written as a functional of the configuration of the membrane, specifically the free energy depends on its curvatures---both mean and Gaussian.

In slender structures such as graphene, there also appears a buckling-like transition from a flat (or rippled, when taking into account thermal fluctuations) state to a buckled state under mechanical load \cite{hanakata_thermal_2021,plummer_buckling_2020}. In addition, scanning tunneling microscopy (STM) experiments have shown that, even in absence of mechanical load, local heating induces a transition from a soft rippled sheet to a hard buckled graphene membrane~\cite{schoelz_graphene_2015,neek-amal_thermal_2014,lindahl_determination_2012,eder_probing_2013}, similar to the Euler buckling transition in structural engineering. This  is somehow unexpected, since the increase of temperature makes the membrane go from a less ordered state (the flat or rippled one) to a more ordered (buckled) state. Due to the strong resemblance between this transition and Euler's buckling under stress, and following the terminology usually employed in previous literature, we also refer to this phenomenon as buckling.
 
The complexity of the interactions in real low-dimensional elastic systems makes it difficult to obtain analytical results, either exactly or with controlled approximations, that improve our understanding of the observed transitions. Therefore, mesoscopic models are relevant, since they contain---in a simplified way---the main ingredients of low-dimensional elastic systems and allow for a detailed analytical approach. Specifically, models based on the coupling between elastic degrees of freedom and Ising spins, termed spin-string and spin-membrane in the one-dimensional and two-dimensional cases, respectively, are able to reproduce in a qualitative way the aforementioned transition. 
Despite their simplicity, these mesoscopic models present a rich behaviour, with a complicated phase diagram that include buckled phases~\cite{ruiz-garcia_stm-driven_2016,ruiz-garcia_bifurcation_2017,ruiz-garcia_ripples_2015}.

Nevertheless, the currently available spin-string and spin-membrane models~\cite{ruiz-garcia_stm-driven_2016,ruiz-garcia_bifurcation_2017,ruiz-garcia_ripples_2015}, although giving an appealing physical picture, are not completely satisfactory from a fundamental point of view. First, some of the microscopic parameters controlling the interactions have to depend on the system size to ensure that the thermodynamic properties have the correct behaviour in the large system size limit, e.g. an intensive transition temperature and an extensive free energy. Second, they are not fully consistent with the classical theory of elasticity~\cite{landau_elasticity_1986}, since rotational symmetry is broken in absence of external forces. 

In this work, we propose the simplest model that keeps the appealing physical picture of the models in Refs.~\cite{ruiz-garcia_stm-driven_2016,ruiz-garcia_bifurcation_2017,ruiz-garcia_ripples_2015} and mends the above mentioned fundamental flaws. We consider a classical elastic lattice (one or two-dimensional) characterised by the out-of-plane deformations and  internal degrees of freedom described by a pseudospin variable~\footnote{We use the term pseudospin to emphasise that it is not the real spin but a two-state variable that mimics in a simple way the range of  values taken by the internal degrees of freedom.}. Elastic deformations are modelled by a harmonic interaction between the atoms that depends on the curvature of the membrane, the repulsive interaction between electrons is modelled by an antiferromagnetic interaction between pseudospins, and the electron-phonon coupling is modelled by the coupling between pseudospins and the membrane curvature. No external stress is acting on the system. In this framework, we will show that the equilibrium state of the system can be completely characterised by a free energy that depends on the membrane curvature---in close analogy with the phenomenological approach based on the Helfrich model~\cite{helfrich_elastic_1973,lipowsky_conformation_1991,wei_bending_2013}. The equilibrium value of the curvature only depends on two parameters: temperature $T$ and  antiferromagnetic coupling constant $J$. 

In this work, we will prove that equilibrium curvature is spatially homogeneous and can be interpreted as the order parameter of the rippled to buckled transition. This is true both for the one-dimensional and the two-dimensional lattices, regardless of the exact geometry of the two-dimensional lattice---as long as it does not contain triangular loops. The phase diagram of the system is built in the $(J,T)$ plane, with an approach that combines analytical and numerical results. This allows us to characterise both the existence and stability of all the possible phases.

In the spin-string model (one-dimensional case), the knowledge of the partition function for the pseudospins make it possible to derive an analytical expression for the free energy for all values of $(J,T)$. Therefrom, many analytical results can be derived. In particular, the transition lines are obtained analytically. In absence of antiferromagnetic coupling, $J=0$, the transition from the rippled, zero-curvature, phase (termed ZC) to the buckled phase (termed B) is second-order, at a certain temperature $T_0$. As the antiferromagnetic coupling is increased, the second-order transition from rippled to buckled moves to lower temperatures and at a certain critical point $(J_c,T_c)$ the transition changes to first order.  Beyond this tricritical point, basically for low enough temperatures in a given range of $J$'s, the behaviour of the system becomes more complex: there appear two buckled phases, one of them locally stable (termed B+) and the other unstable (termed B-). Moreover, there is metastability: the rippled phase ZC is also locally stable---corresponding to a local minimum of the free energy, and which phase, B+ or ZC, gives the absolute minimum of the free energy depends on the value of the antiferromagnetic coupling. This picture is supported by the analytical results that can be worked out close to the second-order bifurcation curve and in the limit of very low temperatures, and also by the numerical construction of the complete phase diagram of the system in the $(J,T)$ plane. 

In the spin-membrane system (two-dimensional case), there is not an analytical closed form of the partition function for the pseudospins. Still, some analytical results can be obtained, specifically in the limits of absence of antiferromagnetic coupling among pseudospins or very low temperatures---for lattices without triangular loops in the latter case.  Despite the increase in the dimensionality of the system, the qualitative picture is similar to the spin-string system. For $J=0$, there appears a  second-order transition from a rippled phase (ZC) to a buckled phase (B). In the limit $T\to 0^+$, the transition is first-order and there is metastability:  again, two buckled phases B$\pm$ emerge, one locally stable and the other unstable, and the zero-curvature phase ZC is also locally stable. In complete analogy with the one-dimensional situation, the relative stability of the B+ and ZC phases depends on the intensity of the antiferromagnetic coupling.

The paper is organised as follows. In Sec.~\ref{sec:model}, we define the one-dimensional model and study its equilibrium behaviour, by deriving a free energy functional from the probability of a given profile. In Sec.~\ref{sec:eq-profiles}, we introduce the Euler-Lagrange equation, which provides us with the equilibrium profile, and define dimensionless variables. In Sec.~\ref{sec:buckling}, we analyse, both analytically and numerically, the emergence of buckled states and put forward the main results of our study. In Sec.~\ref{sec:spin-membrane-honeycomb}, we generalise our one-dimensional model to the two-dimensional case, focusing ourselves on the specific case of a honeycomb lattice---that of actual graphene. We present the main conclusions of our work and perspectives for future research in Sec.~\ref{sec:conclusions}. The
appendices deal with some non-essential technical details that are omitted in the main text.

\section{Rotationally invariant spin-string model}\label{sec:model}

We consider a string on a one-dimensional lattice, with lattice
constant $a$. At each site, $j=0,\ldots,N$, there is a particle of mass $m$ that is characterised
by its transversal displacement $u_{j}$,  and, in
addition, there is a
``pseudospin'' variable $\sigma_j=\pm 1$.  We introduce the following rotationally invariant
energy (or Hamiltonian) for any configuration of the elastic modes
$\bm{u}$, their associated momenta $\bm{p}$, and the pseudospins $\bm{\sigma}$,
\begin{align}
    \mathcal{H}(\bm{u},\bm{p},\bm\sigma)=
    \sum_{j=0}^N &
  \bigg[  \frac{p_j^2}{2m}+
           \frac{k}{2}\left(u_{j+1}-2u_j+u_{j-1}\right)^2 \nonumber \\ 
           - & h
                  \left(u_{j+1}-2u_j+u_{j-1}\right)\sigma_j
                                            + J
  \sigma_{j+1}\sigma_j\bigg].
  \label{eq:energy-rot-inv}
\end{align}
The first term on the rhs is the kinetic energy, the
second one stands for the elastic contribution to the energy---with
elastic constant $k$, the third one involves the interaction between
the elastic displacements and the pseudospins---the parameter $h$
tunes the strength thereof, and the last one represents 
the nearest-neighbour interaction among the
pseudospins---with coupling constant $J$. 
Our Hamiltonian~\eqref{eq:energy-rot-inv} can be analysed for both positive and negative values of the coupling constant $J$. In the case of our concern, the interaction between pseudospins models the repulsive interaction between electrons, as discussed in detail below, and thus we focus on the antiferromagnetic case $J>0$~\footnote{Yet, at some specific points in the paper we briefly consider the situation for ferromagnetic interaction, $J<0$.}.
The pseudospins do
not correspond to actual spin variables, they model in a simple way
internal degrees of freedom---as discussed below.

The above system is a mesoscopic toy model, which tries to resemble in
a simple manner some of the main features of low-dimensional elastic
systems, such as graphene. Therein, the interaction between
the elastic modes and the pseudospins would model the electron-phonon
coupling~\footnote{More specifically, the coupling between the
  transverse elastic modes and out-of-plane electrons, which can be
  thought to be in two states corresponding to the two different sides
  of the plane.}, and the antiferromagnetic interaction between the
pseudospins account for the Coulomb repulsion of
electrons. Therefore, nearest-neighbour out-of-plane
  electrons prefer to be at different sides of the plane.  There are
two main differences with respect to other proposals in the
literature~\cite{ruiz-garcia_stm-driven_2016,ruiz-garcia_bifurcation_2017,ruiz-garcia_ripples_2015}:
(i) the elastic term is proportional to the square of the (discrete)
curvature and not to the square of the (discrete) gradient, and (ii)
the coupling between the elastic modes and the pseudospins also
involves the curvature and not the transverse displacement. 

The key
role played by the curvature of the system, as shown below, makes
our model consistent with the classical theory of
elasticity~\cite{landau_elasticity_1986}---at variance with
previous spin-string (or spin-membrane in the two-dimensional
case)
models~\cite{ruiz-garcia_stm-driven_2016,ruiz-garcia_bifurcation_2017,ruiz-garcia_ripples_2015}. In
particular, a flat profile, $u_{j}\equiv 0$, and a rigid rotation
thereof, $u_{j}=A j$ with constant $A$, have the same
energy in Eq.~\eqref{eq:energy-rot-inv}. In general, any two profiles differing just by a linear function $Aj+B$ has the same energy. The case $A=0$ corresponds to a rigid translation, whereas the case $A\ne 0$ corresponds to a small rotation of angle $A$, $\sin(A) \simeq A$.

Thermal equilibrium of the system at temperature $T$ is described by
the canonical distribution
$P_{\eq}(\bm{u},\bm{p},\bm{\sigma})=\exp\left[-\beta
  \mathcal{H}(\bm{u},\bm{p},\bm\sigma)\right]/Z$,
where $\beta=(k_{B}T)^{-1}$, with $k_{B}$ and $Z$ being the Boltzmann
constant and the partition function, respectively. We are interested in the
equilibrium profiles of the string, so we integrate over $\bm{p}$ and
$\bm{\sigma}$ to get $P_{\eq}(\bm{u})\propto e^{-\beta \calF(\bm{u})}$,
where the free energy of the string $\calF(\bm{u})$ is
\begin{equation}
  \calF(\bm{u})=\sum_{j=0}^N
  \left[\frac{k}{2}\left(u_{j+1}-2u_j+u_{j-1}\right)^2
  \right]-k_{B}T\ln Z_{\intern}(\bm{u}).
\end{equation}
Above, $Z_{\intern}(\bm{u})$ stands for the partial partition function associated
with the internal degrees of freedom, which involves a sum over all
the pseudospin configurations
\begin{equation}
  Z_{\intern}=\sum_{\bm{\sigma}}e^{-\beta\sum_{j=0}^{N}
    \left[-h\left(u_{j+1}-2u_j+u_{j-1}\right)\sigma_j
                                + J \sigma_{j+1}\sigma_j \right] }.
\end{equation}

Now we introduce a continuum limit by assuming that the displacements
$u_{j}$ vary slowly with $j$, i.e., $u_{j}\to u(x)$ with $x=ja$. The
total length of the system is $L=Na$.  In this way,
\begin{equation}\label{eq:u-(j+1)-u-j-u-(j-1)}
  u_{j+1}-2u_{j}+u_{j-1}= a^{2}\chi(x)+O(a^{4}),
\end{equation}
where we have defined the curvature of the profile $\chi(x) \equiv u''(x)$, and
\begin{subequations}
  \begin{align}
    \frac{k}{2}(u_{j+1}-2u_{j}+u_{j-1})^{2}=\frac{k_{0}}{2}\chi^{2},\\
    h (u_{j+1}-2u_{j}+u_{j-1}) \sigma_{j}=h_{0}\chi\sigma_{j}.
\end{align}
\end{subequations}
The continuum limit in the spin variables is skipped willingly, since a marginal sum over all spin configurations will be carried out in brief. Scaled parameters for the elastic constant and the
strength of the pseudospin-elastic interaction have been introduced as
\begin{equation}\label{eq:h0-k0-scalings}
  k_{0}=k a^{4}, \quad   h_{0}=h a^{2}.
\end{equation}
Also, terms with higher powers of $a$ in Eq.~\eqref{eq:u-(j+1)-u-j-u-(j-1)} have been omitted, since they involve higher-order derivatives of $u$~\footnote{Therefore, these terms are expected to be subdominant when $u_{j}$ varies slowly with $j$. The (small) corrections introduced thereby could be incorporated into our theoretical framework in a perturbative way.}.

In the continuum limit, the equilibrium probability of a certain
string profile $u(x)$ becomes a functional thereof, specifically
\begin{equation}
  P_{\eq}[u]\propto e^{-\beta\calF[u]}, \quad \calF[u]=n\int_{0}^{L}
  dx f(\chi),
\end{equation}
where $n\equiv N/L$ is the number density, and the free energy density (per particle) $f$ is given by
\begin{subequations}\label{eq:free-energy-density}
\begin{align}
  f(\chi)=&\,\frac{k_{0}\chi^2}{2}
  -\beta^{-1} \ln\zeta_{\intern}^{\onedim} (\beta h_0 \chi, \beta J), \\
  \zeta_{\intern}^{\onedim}(\beta h_0 \chi, \beta J ) \equiv& \, e^{-\beta J }
  \cosh\left(\beta h_0 \chi  \right) \nonumber \\
  &+e^{\beta J }
                   \sqrt{1+e^{-4\beta J}\sinh^{2}\left( \beta h_0 \chi \right)},                 \label{eq:one-particle-zeta}
\end{align}
\end{subequations}
Our notation explicitly tells us that $f$ only depends on the profile
through its curvature $\chi$. The second term in $f(\chi)$ comes from
the sum of the spin variables, $\zeta_{\intern}^{\onedim}$ is the
one-particle partition function of a one-dimensional 
Ising chain with coupling $J$ and external field $h_{0}\chi$---e.g.~see
Ref.~\cite{feynman_statistical_1996}.

In contrast to previous spin-string and spin-membrane models~\cite{ruiz-garcia_stm-driven_2016,ruiz-garcia_bifurcation_2017,ruiz-garcia_ripples_2015}, the continuum limit consider here does not involve a large system size limit $N\gg 1$ (thermodynamic limit). This approach to the continuum limit is more consistent from a physical point of view. More specifically, the microscopic parameters in the Hamiltonian, such as $h$ and $k$ do not scale with the system size---thus decoupling the continuum and the thermodynamic limits.

\section{Equilibrium profiles}\label{sec:eq-profiles}

The equilibrium profiles $u_{\eq}(x)$ are those that maximise the
probability distribution $P_{\eq}[u]$ or, equivalently, those that
minimise the free energy functional $\calF[u]$. Therefore, we consider the first
variation of the free energy functional upon the change $u\to u+\delta
u$, 
\begin{align}
         \delta F[u]=&\int_0^L \!\!\! d{x}\,  \delta u\, \frac{d^{2}}{dx^{2}}\!\pr{\pdv{f}{\chi}}
   \!\!+\! \corch{\delta u'\,\pdv{f}{\chi}-\delta u\, \dv{}{x}\!
                          \pr{\pdv{f}{\chi}} }_0^L. 
     \label{eq:free-energy-var}
\end{align}
The expression for $\delta F[u]$ is more involved than
in the usual case where the integrand only depends on the first
derivative, but the integral and the boundary terms must vanish
separately nonetheless~\cite{gelfand_calculus_2000,lanczos_variational_1970}.\footnote{For instance, for free boundary conditions, the variation $\delta u(x)$ for $x\in(0,1)$ and
$\{\delta u(0),\delta u(1),\delta u'(0),\delta u'(1)\}$ at the
boundaries are independent.} Considering the vanishing of the integral term, one obtains the Euler-Lagrange
equation
\begin{equation}
  \left.\dv[2]{}{x}\!\pr{\pdv{f}{\chi}}\right|_{\eq}=0 \implies  \left.\pdv{f}{\chi}\right|_{\eq}=Ax+B,
\end{equation}
where $A$ and $B$ are arbitrary constants, to be determined by
imposing the boundary conditions. The boundary conditions depend on
the physical situation: here, for the sake of simplicity we are going
to consider that the ends of the string are fixed, so that
\begin{equation}\label{eq:fixed-ends-bc}
  u(0)=u(L)=0,
\end{equation}
but the value of $u'$ at the boundaries is free---the
so-called supported boundary
conditions~\cite{landau_elasticity_1986}. Therefore, $\pdv*{f}{\chi}$
must vanish at both boundaries for the equilibrium profile, which entails $A=B=0$.

The
equilibrium profile is thus provided by $\left.\pdv{f}{\chi}\right|_{\eq}\!\!\!=0$, i.e.
\begin{align}\label{eq:profile-eq}
k_{0}\chi_{\eq}&=\left. k_B T \pdv{\ln\zeta_{\intern}^{\onedim} 
  }{\chi}\right|_{\chi=\chi_\eq}\!\!\! \nonumber \\ 
  &=h_{0}\frac{e^{-\frac{2J}{k_B T}}\sinh{\left(\frac{h_0 \chi_{\eq}}{k_B T}\right)}}{\sqrt{1+e^{-\frac{4J}{k_B T}}\sinh^2{\left(\frac{h_0 \chi_{\eq}}{k_B T}\right)}}},
\end{align}
with the boundary conditions
\eqref{eq:fixed-ends-bc}. Equation~\eqref{eq:profile-eq} is a transcendental equation for the equilibrium curvature $\chi_{\eq}$, the
solution of which can be written as
$\chi_{\eq}=\chi_{\eq}(T,J,k_{0},h_{0})$.\footnote{Zero curvature profiles, i.e. linear profiles with $\chi=0$, always solve Eq.~\eqref{eq:profile-eq}.} The equilibrium curvature is constant, independent of position, and so is $f_{\eq}=f(\chi_{\eq})$. Therefore, the equilibrium free energy is $\calF_{\eq}=n Lf_{\eq}=Nf_{\eq}$, i.e. it is extensive---note that the constant $\chi_{\eq}$ is an intensive quantity, independent of system size.

If $\chi_{\eq}$ is a solution of Eq.~\eqref{eq:profile-eq}, then
$-\chi_{\eq}$ is also a solution. This is a consequence of the free
energy density being an even function of $\chi$, and it is clear on a
physical basis: the specular reflection of an equilibrium profile with
respect to the $x$ axis is also an equilibrium profile, since there is
no external field. Therefore, without loss of generality, we restrict
ourselves to solutions with positive curvature in the remainder of this work.

To better characterise the equilibrium curvature $\chi_{\eq}$, it is
adequate to introduce dimensionless variables: this allows us to
identify the natural units of length and energy---or temperature---in
our model system, and elucidate the dimensionless combinations of the
system parameters on which the curvature
depends. Equation~\eqref{eq:profile-eq} tells us the characteristic
length $\ell_{0}$, which in turn determines the characteristic
temperature $T_{0}$,
\begin{equation}\label{eq:l0-and-T0}
  \ell_{0}=\frac{k_{0}}{h_{0}}=\frac{k}{h}a^2, \qquad T_{0}=\frac{h_{0}^{2}}{k_{B}k_{0}}=\frac{h^{2}}{k_{B}k}.
\end{equation}
Then we define dimensionless displacement $u^{*}$, position
$x^{*}$, curvature $\chi^{*}$, temperature $T^{*}$, and coupling constant $J^{*}$ by
\begin{equation}\label{eq:dimensionless-variables}
  u^{*}=\frac{u}{\ell_{0}}, \,\,  x^{*}=\frac{x}{\ell_{0}}, \,\,
  \chi^{*}=\ell_{0}\chi, \,\,
  T^{*}=\frac{T}{T_{0}}, \,\, J^{*}=\frac{J}{k_{B}T_{0}}.
\end{equation}
Consistently, we define $\beta^*=k_B T_0 \beta=1/T^*$,  $L^{*}=L/\ell_{0}$, $n^*=n\ell_0=N/L^*$, and the dimensionless free energy $\calF^{*}\equiv\calF/k_{B}T_{0}$. Then, we can write
\begin{subequations}
\begin{align}
  \calF^{*}[u^{*}](J^{*},T^{*})&=n^*\int_{0}^{L^{*}}  dx^{*}
  f^{*}(\chi^{*};J^{*},T^{*}), \\
 f^{*}(\chi^{*};J^{*},T^{*})&\equiv \frac{1}{2}{\chi^{*}}^{2}-T^{*}\ln
                              \zeta_{\intern}^{\onedim}(\beta^* \chi^{*},\beta^* \! J^{*}),
                              \label{eq:non-dim-free-energ-dens}
\end{align}
\end{subequations}
where $f^{*}$ is the dimensionless free energy density and $\zeta_{\intern}^{\onedim}$ is the same one-particle partition function for the pseudospins than that in Eq.~\eqref{eq:one-particle-zeta}.\footnote{Note that $\beta h_0 \chi = \beta^* \chi^*$ and $\beta J = \beta^* J^*$.}

Our use of dimensionless variables simplifies the theoretical analysis, by reducing the number of parameters---and identifying the relevant combinations thereof. This course of action aligns well with our main goal, which is the explanation of the experimentally observed buckling phenomenon with a simple model. A quantitative comparison of the results predicted by our model with experimental data of a specific system is beyond the scope of this work, and in any case would involve fitting the model parameters to the experimental data.

In dimensionless variables, Eq.~\eqref{eq:profile-eq} is rewritten as $\left.\pdv{f^{*}}{\chi^{*}}\right|_{\eq}=0$, i.e. 
\begin{subequations}\label{eq:profile-eq-nondim}
\begin{align}
   \chi_{\eq}^{*}&=\mu^*(\beta^* \chi_\eq^*,\beta^* \!J^*), \\
  \mu^*(\beta^* \chi^*,\beta^* \!J^*)&\equiv \pdv{\ln\zeta_{\intern}^{\onedim}}{(\beta^* \chi^*)}=\!\frac{e^{-{2\beta^*\!J^{*}}}
    \sinh{\left(\beta^* \chi^*\right)}}
  {\sqrt{1\!+\!e^{-4\beta^*\!J^{*}}\sinh^2{\left(\beta^* \chi^*\right)}}},
\end{align}
\end{subequations}
where $\mu^*$ is the local magnetisation of the pseudospins.
Equation~\eqref{eq:profile-eq-nondim} implies that $\chi_{\eq}^{*}$ is a certain function of $J^*$ and $T^*$, $\chi_{\eq}^{*}=\chi_{\eq}^{*}(J^{*},T^{*})$. Therefore, it is the dimensionless pseudospin-elastic coupling $J^{*}$
and the dimensionless temperature $T^{*}$ that control the emergence
of buckled profiles with non-zero curvature. To keep our notation simple, we do not explicitly state that both $J^*$ and $T^*$ depend on the microscopic parameters of the model through the characteristic temperature $T_0$, defined in Eq.~\eqref{eq:l0-and-T0}, employed to make energy dimensionless. We recall that the equilibrium
curvature is constant, independent of position, and an intensive quantity---and so is the
free energy density. Therefore, 
\begin{equation}\label{eq:free-energy-equil}
  \calF_{\eq}^{*}(J^{*},T^{*})\equiv \calF^{*}[u_{\eq}^{*}](J^{*},T^{*})=N
  f^{*}(\chi_{\eq}^{*};J^{*},T^{*}),
\end{equation}
the equilibrium free energy is extensive. 

For the supported boundary
conditions~\eqref{eq:fixed-ends-bc} we are considering, there is a unique smooth equilibrium profile
given by
\begin{equation}\label{eq:profile-sol-supported}
  u_{\eq}^{*}(x^{*};J^{*},T^{*})=
  \frac{\chi_{\eq}^{*}(J^{*},T^{*})}{2}x^{*}(x^{*}-L^{*}),
\end{equation}
For completely free boundary conditions, i.e. when neither the
displacement nor its derivative is fixed at the boundary, there are
multiple equilibrium profiles that are obtained by adding $Cx^{*}+D$ to
$u_{\eq}^{*}$ in Eq.~\eqref{eq:profile-sol-supported}, with $C$ and $D$
being arbitrary constants---the boundary condition
$\dv*{(\pdv*{f^{*}}{\chi^{*}})}{x^{*}}|_{x^{*}=0,L^{*}}=0$ is
fulfilled for all $(C,D)$. In particular, the flat profile, $u^{*}=0$,
and any transversal shift plus rotation thereof, $u^{*}=Cx^{*}+D$, are both
possible equilibrium profiles for free boundary conditions.

Since the free energy density only depends on the curvature, and the
equilibrium curvature is the same for both supported and completely
free boundary conditions, the analysis of buckled states that follows
is valid for both supported and free boundary conditions. Then, for
the sake of concreteness and without loss of generality, we will
employ the supported boundary condition
expression~\eqref{eq:profile-sol-supported} for the equilibrium
profile---omitting the additional linear contribution $Cx^{*}+D$ for the
free case.

In the remainder of the paper, we employ dimensionless
variables---dropping the asterisks in them not to clutter our formulae.

\section{Buckled states}\label{sec:buckling}

In this section, we analyse the emergence of buckled states, i.e. with
non-zero curvature, in our model system. First, we show how buckled
profiles bifurcate from the zero-curvature solution $\chi_{\eq}^{\zerocurv}=0$ of
Eq.~\eqref{eq:profile-eq-nondim}. Second, we consider the low-temperature
limit $T\ll 1$, where the equilibrium profiles can be exactly
calculated. Lastly, we analyse the phase diagram of the model by numerically solving  Eq.~\eqref{eq:profile-eq-nondim}.

\subsection{Bifurcation from the zero-curvature solution}\label{sec:bifurc-from-flat-sol}

To start with, we expand the free energy density around the
zero-curvature profile, for which
\begin{equation}
  f_{\zerocurv}(J,T)\equiv f(\chi=0;J,T)=-T\ln\corch{2\cosh(J/T)},
\end{equation}
i.e. we consider the difference of the free energy density from the
zero-curvature value,
\begin{align}
  \Delta f(\chi;J,T)&\equiv f(\chi;J,T)-f_{\zerocurv}(J,T)\nonumber \\
  &=\frac{1}{2}\chi^{2}-T\ln\corch{
  \frac{\zeta_{\intern}^{\onedim}( \chi/T,J/T)}{2\cosh(J/T)}}.
\label{eq:Deltaf-def}
\end{align}
The stability of the zero-curvature profile is determined by the sign
of the second-derivative of the free energy density at $\chi=0$,
\begin{equation}\label{eq:f2}
  f_2(J,T)\equiv \left.\pdv[2]{f}{\chi}\right|_{\chi=0}=
  1-\frac{e^{-2J/T}}{T}. 
\end{equation}
The change of stability takes place at the line over which $f_{2}$
vanishes, i.e.
\begin{equation}\label{eq:bifurc-curve}
  J=J_{b}(T)=-\frac{1}{2}T\ln T,
\end{equation}
which is called the bifurcation curve. This line separates the plane
$(J,T)$ into two regions: (i) a region with $f_{2}>0$, comprising two subregions I and III  where the
zero-curvature solution is stable and (ii) region II with $f_{2}<0$,
where the zero-curvature solution becomes unstable and other stable
solution exist. Note that for $J=0$, we have
that $f_{2}<0$ for $T<1$: the unit of temperature $T_{0}$ is thus the
transition temperature in the absence of antiferromagnetic coupling
among the pseudospins. 

The deviation of the free energy density from the zero-curvature
value is provided by the Taylor expansion
\begin{align}
  \Delta f(\chi;J,T)=&\frac{f_2(J,T)}{2} \chi^2 +
\frac{f_4(J,T)}{4!} \chi^4 & \nonumber \\ &+
  \frac{f_6(J,T)}{6!} \chi^6+O\left(\chi^8\right),
  \label{eq:free-energy-expansion}
\end{align}
where 
\begin{subequations} \label{eq:f4-f6}
\begin{align}
  f_4(J,T)=&\frac{e^{-6J/T}}{T^3}\!\left(3-e^{4J/T}\right),  \\ 
  f_6(J,T)=&-\frac{e^{-10J/T}}{T^5} 
  \!\left(e^{8J/T}-30e^{4J/T}+45\right).
\end{align}
\end{subequations}
\begin{figure} \centering
\includegraphics[width=3.25in]{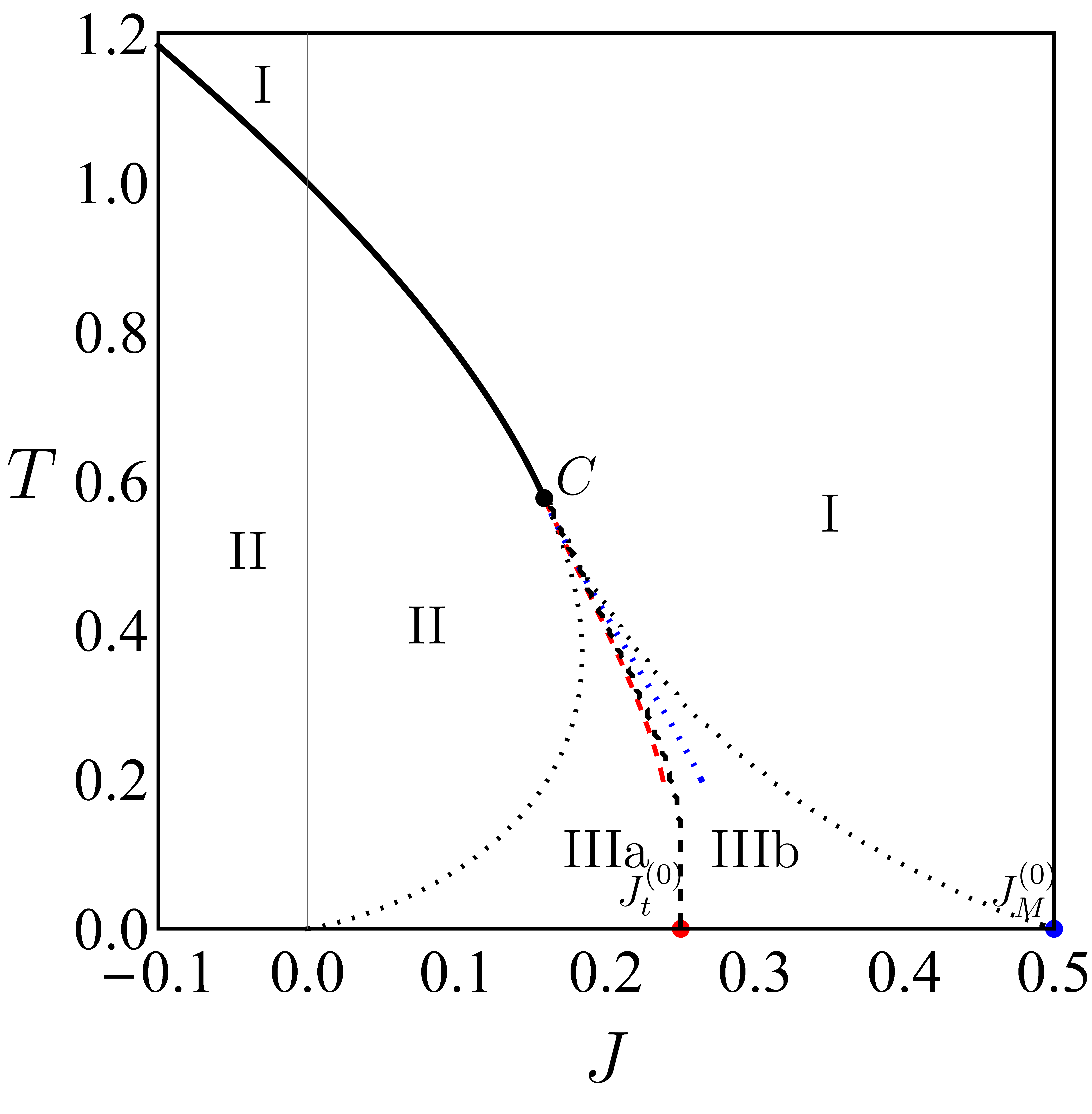}
\caption{Phase diagram of the rotationally invariant spin-string. The plane $(J,T)$ is divided into several regions, characterised by the existing phases and their stability. In region I, only the ZC phase exists. In region II, phase ZC is unstable, and there appears a stable buckled phase B. In region III, three phases coexist: ZC, and two buckled phases, B+ and B-; the latter being unstable, whereas ZC and B+ are locally stable. Phase B+ is the most stable one in IIIa, with ZC being metastable, whereas the roles are reversed in IIIb. Point $C=(J_{c},T_{c})$ is the tricritical point. Above it, the phase transition is second-order; below, it is first-order.  The curve $J_b(T)$, given by Eq.~\eqref{eq:bifurc-curve}, is a bifurcation curve. For $T>T_c$, it defines the already mentioned second-order phase transition (black solid line). For $T<T_{c}$, it demarcates the change from region II to region III (leftmost dotted line). The curve $J_{M}(T)$ (rightmost dotted line), theoretically predicted by Eq.~\eqref{eq:limit-metastab-curve-main} close to the tricritical point (blue dotted line), demarcates the change from region III to region I. The curve $J_{t}(T)$ (dashed line), theoretically predicted by Eq.~\eqref{eq:first-order-curve-main} close to the tricritical point (red dashed line), is the first-order transition line: over it, phases B+ and ZC have the same free energy. Also plotted are the theoretical low-temperature limiting values $J_{M}^{(0)}$ and $J_{t}^{(0)}$. On the one hand, the predictions for the bifurcation curve and for $\left(J_t^{(0)},J_M^{(0)}\right)$ are exact. On the other hand, the asymptotic theoretical predictions for the curves $J_t(T)$ and $J_M(T)$ show an excellent agreement with the numerical results, within their range of validity. See Sec.~\ref{sec:num-analys} for further details in the numerical determination of the curves $J_t(T)$ and $J_M(T)$.}
\label{kappavstheta3}
\end{figure}

The phase diagram of the system is presented in
Fig.~\ref{kappavstheta3}, paying special attention to the demarcation of the different regions and subregions where both the number of equilibrium solutions and its stability change. This figure summarises the main results of the equilibrium and stability analyses  carried out in detail in Appendix~\ref{sec:bifurc-theory}. Therefrom, we highlight here the tricritical point,
\begin{equation}
T_c=\frac{1}{\sqrt{3}} , \quad J_c=J_b(T_c)= \frac{\ln 3}{4\sqrt{3}}
\end{equation}
and the asymptotic theoretical prediction for the curves $J_M$ ($J_t$) separating regions I and IIIb (IIIb and IIIa),
\begin{subequations}
    \begin{align}
\label{eq:limit-metastab-curve-main}   J_M(T)=J_b(T) + \frac{5}{12}\sqrt{3}(T-T_c)^2 \\
 \label{eq:first-order-curve-main}  J_t(T)=J_b(T) + \frac{5}{16}\sqrt{3}(T-T_c)^2 .       
    \end{align}
\end{subequations}

\subsection{Low-temperature limit}\label{sec:low-temp-limit}

Now we turn our attention to the low-temperature limit $T\ll 1$. By defining $e_{\gs}^{\onedim}$ as the energy per site in the ground state of the one-dimensional pseudospin system with external field $\chi$ and antiferromagnetic coupling $J$, we can write
\begin{equation}\label{eq:egs-def}
   \lim_{T\to 0^+}-T\ln\zeta_{\intern}^{\onedim}=e_{\gs}^{\onedim}.
\end{equation}
The ground state energy of the pseudospin system is
\begin{equation}\label{eq:egs-expression-1d}
    e_{\gs}^{\onedim}=-\pr{\left|\chi\right|-2J}H\pr{\left|\chi\right|-2J}-J,
\end{equation}
in which  $H(x)$ is Heaviside's step function. Equation~\eqref{eq:egs-expression-1d} is readily understood on a physical basis: for small $\left|\chi\right|$, the antiferromagnetic coupling wins and the ground state corresponds to antiferromagnetic ordering, $e_{\gs}^{\onedim}=-J$, whereas for large $\left|\chi\right|$, the external field wins and the ground state corresponds to all spins aligned with the curvature, $e_{\gs}^{\onedim}=-\left|\chi\right|+J$. These two expressions for $e_{\gs}^{\onedim}$ become equal at $\left|\chi\right|=2J$, which marks the crossover between them both. Of course, $e_{\gs}^{\onedim}$ can also be derived by taking the limit $T\to 0^+$ in Eq.~\eqref{eq:egs-def}, bringing to bear Eq.~\eqref{eq:one-particle-zeta}
for $\zeta_{\intern}^{\onedim}$.

The above discussion entails that the free energy density \eqref{eq:non-dim-free-energ-dens} simplifies to
\begin{equation}\label{eq:free-energ-dens-low-T}
    f(\chi;J,T)\sim\frac{\chi^2}{2}
    -\pr{\left|\chi\right|-2J}H\pr{\left|\chi\right|-2J}-J, \quad T\ll 1.
\end{equation}
Also, Eq.~\eqref{eq:profile-eq-nondim} for the equilibrium curvature converts to
\begin{align}\label{eq:EL-low-T}
    \abs{\chi_{\eq}}=H(\abs{\chi_{\eq}}-2J), \quad
  T\ll 1,
\end{align}
Both in Eqs.~\eqref{eq:free-energ-dens-low-T} and \eqref{eq:EL-low-T},
Heaviside's step function must be understood in a ``physical way'',
since it appears here as the low-temperature limit of the rhs of
Eq.~\eqref{eq:profile-eq-nondim}. Therefore, $H(x)$ comprises three
strokes: two flat strokes, equal to $0$ and $1$ for $x<0$ and $x>0$,
respectively, and a vertical stroke at $x=0$, which joins the previous
two. Logically, Eq.~\eqref{eq:EL-low-T} preserves the symmetry
$\chi_{\eq}\to-\chi_{\eq}$: we recall that we are restricting
ourselves to positive curvatures, so $\abs{\chi_{\eq}}\to\chi_{\eq}$ in
the following. 

\begin{figure}
    \centering
    \includegraphics[width=3.25in]{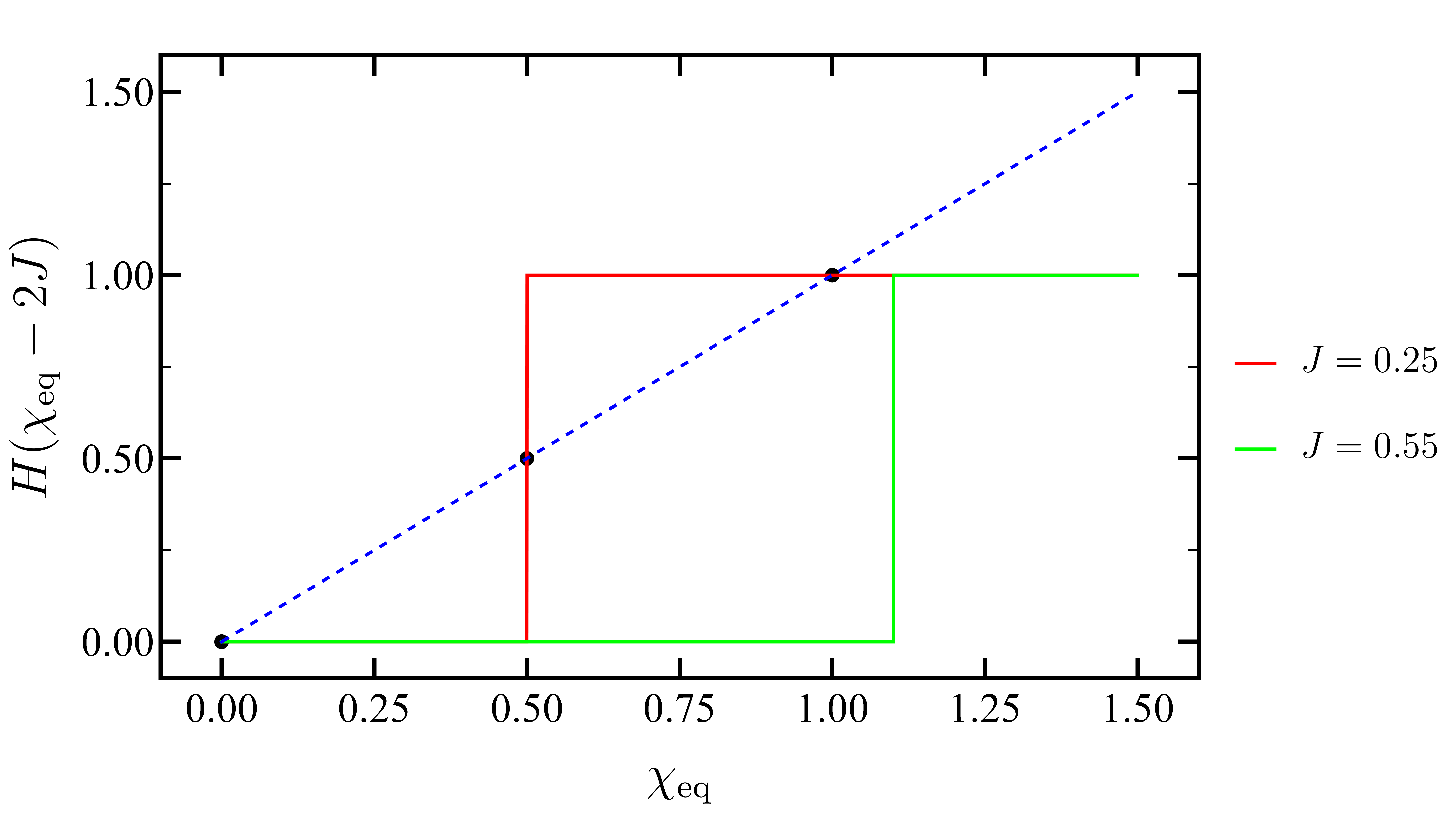}
    \caption{Graphical method for solving Eq.~\eqref{eq:EL-low-T}. For each $J$, the
      intersection points of the functions $H(\chi_{\eq}-2J)$ (solid lines) and $\chi_{\eq}$ (dashed blue line) give the equilibrium
      curvature  in the low-temperature limit. For $J > J_{M}^{(0)}=1/2$,
      for instance $J=0.55$ (green solid line), the only solution is
      $\chi_{\eq}=0$, i.e. the ZC phase. For $J\leq J_{M}^{(0)}$, for instance
      $J=0.25$ (red solid line), there appear three solutions: in
      addition to the ZC phase corresponding to $\chi_{\eq}=0$, we
      have two buckled phases
      corresponding to $\chi_{\eq}=2J$ (B- phase) and $\chi_{\eq}=1$ (B+ phase).}
    \label{fig:LowTemperatures}
\end{figure}

In order to study the solutions of Eq.~\eqref{eq:EL-low-T}, a
graphical method is useful. In Fig.~\ref{fig:LowTemperatures}, we plot
the functions on the lhs and the rhs of Eq.~\eqref{eq:EL-low-T} and
look for the intersection points of both functions. It is clearly seen
that $\chi_{\eq}=0$ is always a solution, for all $J$: the phase ZC
survives in the low-temperature limit. In addition, there appear two
buckled solutions for $2J\leq 1$, i.e. $J\leq J_{M}^{(0)}= 1/2$:
specifically, $\chi_{\eq}=2J$ and $\chi_{\eq}=1$. Both solutions
coalesce at $J_{M}^{(0)}$: stronger antiferromagnetic interaction destroys
the buckled phases, in which the magnetisation of the pseudospins is different from zero---a signature of ferromagnetic ordering.

Let us calculate the free energy density of each of the phases, in
order to identify them. First, the free energy density of the ZC phase
is given by $f_{\zerocurv}(J,T)\sim -J$, for $T\ll 1$, so that
\begin{align}
  \Delta f(\chi;J,T)&=
  f(\chi;J,T)-f_{\zerocurv}(J,T) \nonumber \\ 
  &\sim
  \frac{\chi^2}{2}
  -\pr{\left|\chi\right|-2J}H\pr{\left|\chi\right|-2J},
  \quad T\ll 1.
\end{align}
Therefore, one obtains
\begin{subequations}
\begin{align}
    \Delta f(\chi_{\eq}=&2J; J,T\to0)\sim 2J^{2}, \\
    \Delta f(\chi_{\eq}=&1;J,T\to0)\sim -\frac{1}{2}+2J.
\end{align}
\end{subequations}
On the one hand, the phase with $\chi_{\eq}=2J$ has always a free
energy that is larger than that of the ZC phase---recall that the total free energy is proportional to its density as stated by 
Eq.~\eqref{eq:free-energy-equil}. On the other hand, $\Delta f(\chi_{\eq}=1;J,T\to0)$ changes sign at $J=J_{t}^{(0)}=1/4$: the phase
with $\chi_{\eq}=1$ has a free energy that is smaller (larger) than
that of the ZC phase for $0\leq J<J_{t}^{(0)}$
$\left(J_{t}^{(0)}<J\leq J_{M}^{(0)}\right)$. Therefore, the solution with
$\chi_{\eq}=1$ is expected to correspond to the low-temperature limit
of phase B+, whereas the solution with $\chi_{\eq}=2J$ should
correspond to the low-temperature limit of  phase B-. Also, the
points $\left(J_{t}^{(0)},0\right)$ and $\left(J_{M}^{(0)},0\right)$ should correspond to
the endpoints of the first-order line $J_{t}(T)$ and the right border line
of region III $J_{M}(T)$, respectively---we have analytical
predictions for these lines, them being rigorously valid just in the vicinity of the tricritical point,
Eqs.~\eqref{eq:first-order-curve-main} and
\eqref{eq:limit-metastab-curve-main}. These
expectations are confirmed by the numerical analysis below.

\subsection{Numerical analysis}\label{sec:num-analys}

Here, we perform a numerical analysis of the phase diagram of our model system. 
We construct a
200$\times$200 mesh in the $(J,T)$ plane and find $\chi_{\eq}$, by numerically solving
Eq.~\eqref{eq:profile-eq-nondim} at
each point of this mesh. At some points $(J,T)$, more than one
solution is found: note that this is expected, since in certain
regions of the plane $(J,T)$ several phases coexist. This analysis
allows us to build a numerical phase diagram for our system and
compare it with our analytical predictions above---depicted in
Fig.~\ref{kappavstheta3}.

\begin{figure}
  \centering
  \includegraphics[width=3.25in]{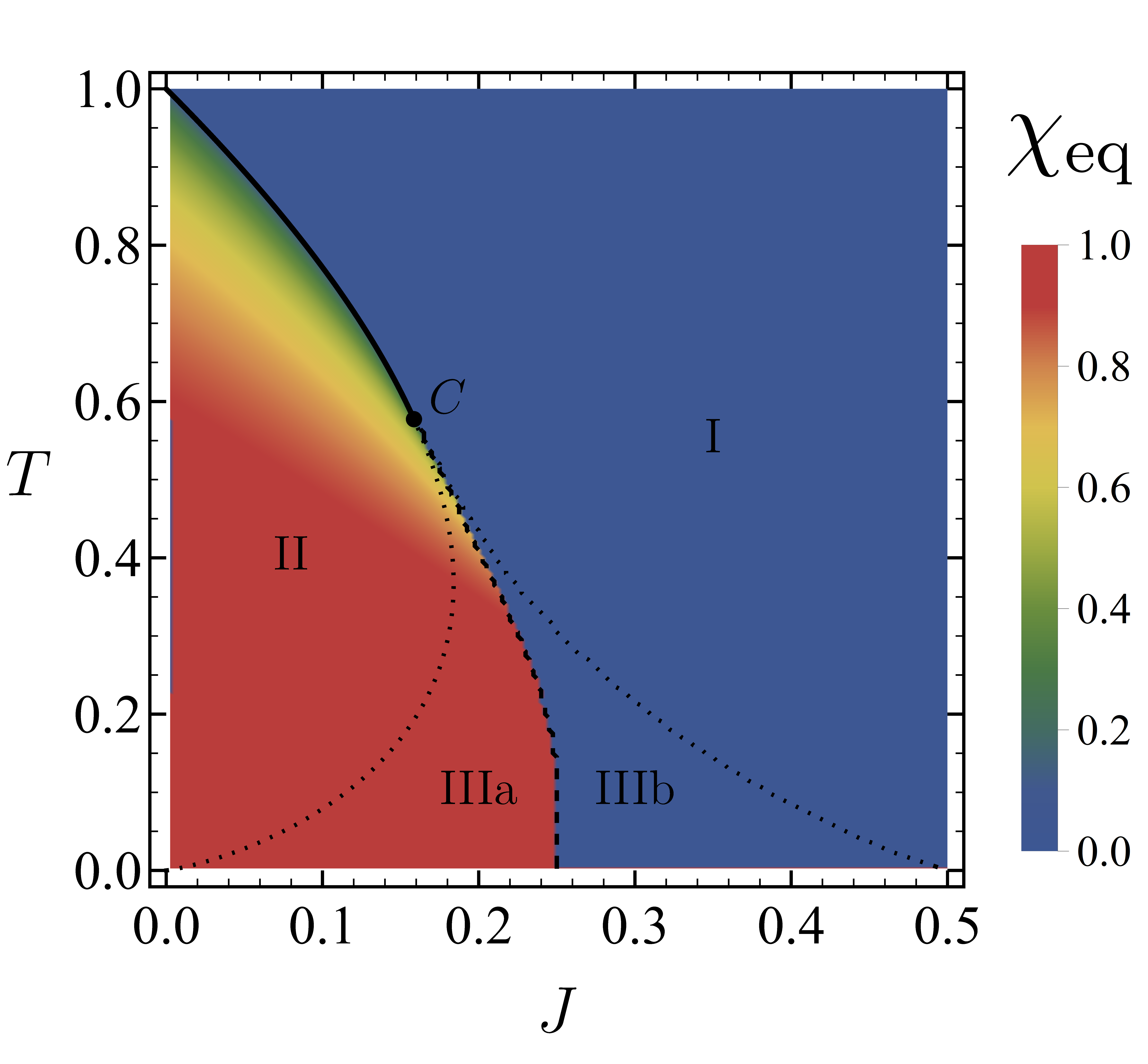}
  \caption{Curvature of the most stable phase (numerically obtained). The labelling of the
    different regions and the code of the different lines is the same
    as in Fig.~\ref{kappavstheta3}. The change of order of the
    transition, from second-order (solid line) above $C$ to first-order (dashed line) below, is
    clearly observed. }
  \label{fig:chieq-num}
\end{figure}

Figure~\ref{fig:chieq-num} shows the values of $\chi_{\eq}$ for the
thermodynamically stable phase---the most stable one when there is
coexistence. Therein, it is clearly observed that the transition
changes from second-order to first-order at the tricritical
point. Above the tricritical point $C$, $\chi_{\eq}$ changes continously
from zero to non-zero values at the bifurcation curve
$J_{b}(T)$ (solid line). Below the tricritical point, $\chi_{\eq}$ changes
discontinuously from zero to non-zero values at the first-order line
$J_{t}(T)$ (dashed line). Recall that the curvature plays the role of the order
parameter in our model. In fact, $\chi_{\eq}$ equals the magnetisation
of the pseudospins, which is given by the rhs of
Eq.~\eqref{eq:profile-eq-nondim}.

\begin{figure*}
    \centering  
    \includegraphics[width=6in]{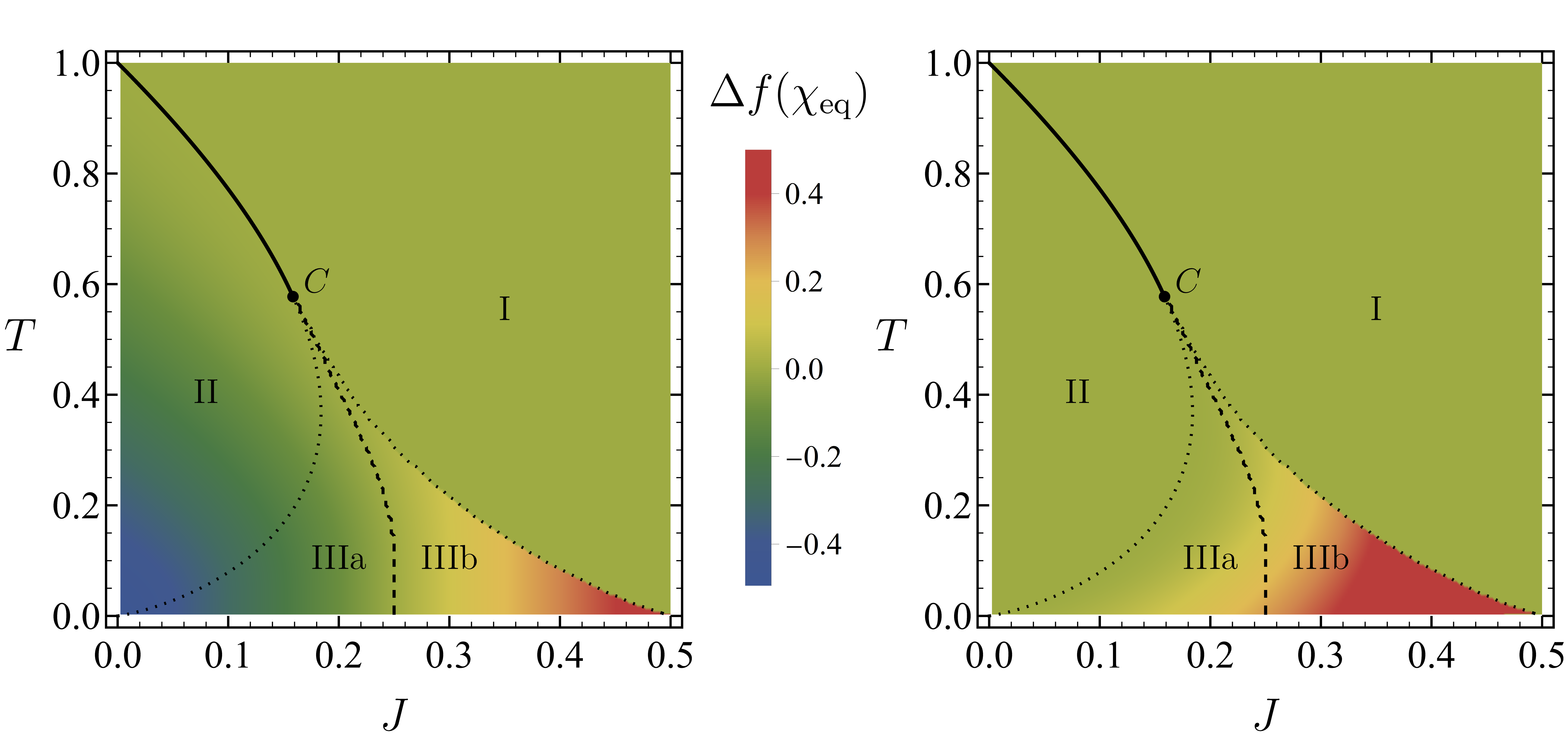} 
    \caption{Free energy density difference $\Delta f(\chi_{\eq};J,T)$ of the buckled states
      with respect to the ZC phase (numerically obtained). The labelling of the different
      regions and the code of the different lines is the same as in
       Figs.~\ref{kappavstheta3} and Fig.~\ref{fig:chieq-num}. On the left panel, we show a
      density plot for the (at least locally) stable buckled phase: B
      in region II and B+ in region III. On the right panel, we show
      the density plot for the unstable buckled phase B-, which only
      exists in region III. 
      }
    \label{fig:phase-diagram_num_200x200}
\end{figure*} 
By inserting the numerical solution for $\chi_{\eq}$ into
Eq.~\eqref{eq:non-dim-free-energ-dens}, we get the values of the free
energy density over the considered mesh.  We present our results in
Fig.~\ref{fig:phase-diagram_num_200x200}, where we plot the
difference of the free energy density with respect to the ZC
profile, $\Delta f(\chi_{\eq};J,T)$, as defined in
Eq.~\eqref{eq:Deltaf-def}. In the left panel, the value for the stable
buckled phase, wherever it exists, is shown---in region I, within
there is no buckled phase, a constant value zero is
plotted. Therefore, phase B is shown in region II, inside the
bifurcation curve \eqref{eq:bifurc-curve} (solid line above the tricritical point $C$, leftmost dotted line below it), and phase B+
is plotted in region III. Region III is demarcated by the bifurcation
curve, the segment of the $J$-axis between the origin and the point
$\left(J_{M}^{(0)},0\right)$, and the curve $J_{M}(T)$ that marks the limit of
existence of the phases B$\pm$ (rightmost dotted line).  In the right panel,
$\Delta f(\chi_{\eq};J,T)$ is plotted for the unstable phase B-, which
only exists in region III---consistently, in regions I and II, a
constant value zero is plotted. The three phases, ZC, B+, and B-
coexist in region III, which is divided into two subregions by the
first-order line $J_{t}(T)$ (dashed line): IIIa, where
$\Delta f(\chi_{\eq}^{\BuckledStable};J,T)<0$ and the most stable
phase is B+, being ZC metastable; and IIIb, where the roles are
reversed, the most stable phase is ZC, being B+ metastable.

\section{Spin-membrane model on a honeycomb lattice}\label{sec:spin-membrane-honeycomb}

Now we move to a two-dimensional system, i.e. a spin-membrane
model. Specifically, we consider the system on a honeycomb lattice,
such as that of graphene. There is a particle---carbon atom if
thinking of graphene---at each node $(i,j)$ in a certain two-dimensional region $\Omega$.
The transverse displacement with respect to the
plane of each particle is denoted by $u_{ij}$. Analogously to the
one-dimensional spin-string analysed before, displacements parallel to
the plane are not taken into consideration. Moreover, at each site
$(i,j)$ we have a pseudospin $\sigma_{ij}=\pm 1$, which models in a
simple way other degrees of freedom---for example, the out-of-plane
electron that is not bonded.

\begin{figure}
    \centering
    \includegraphics[width=3.25in]{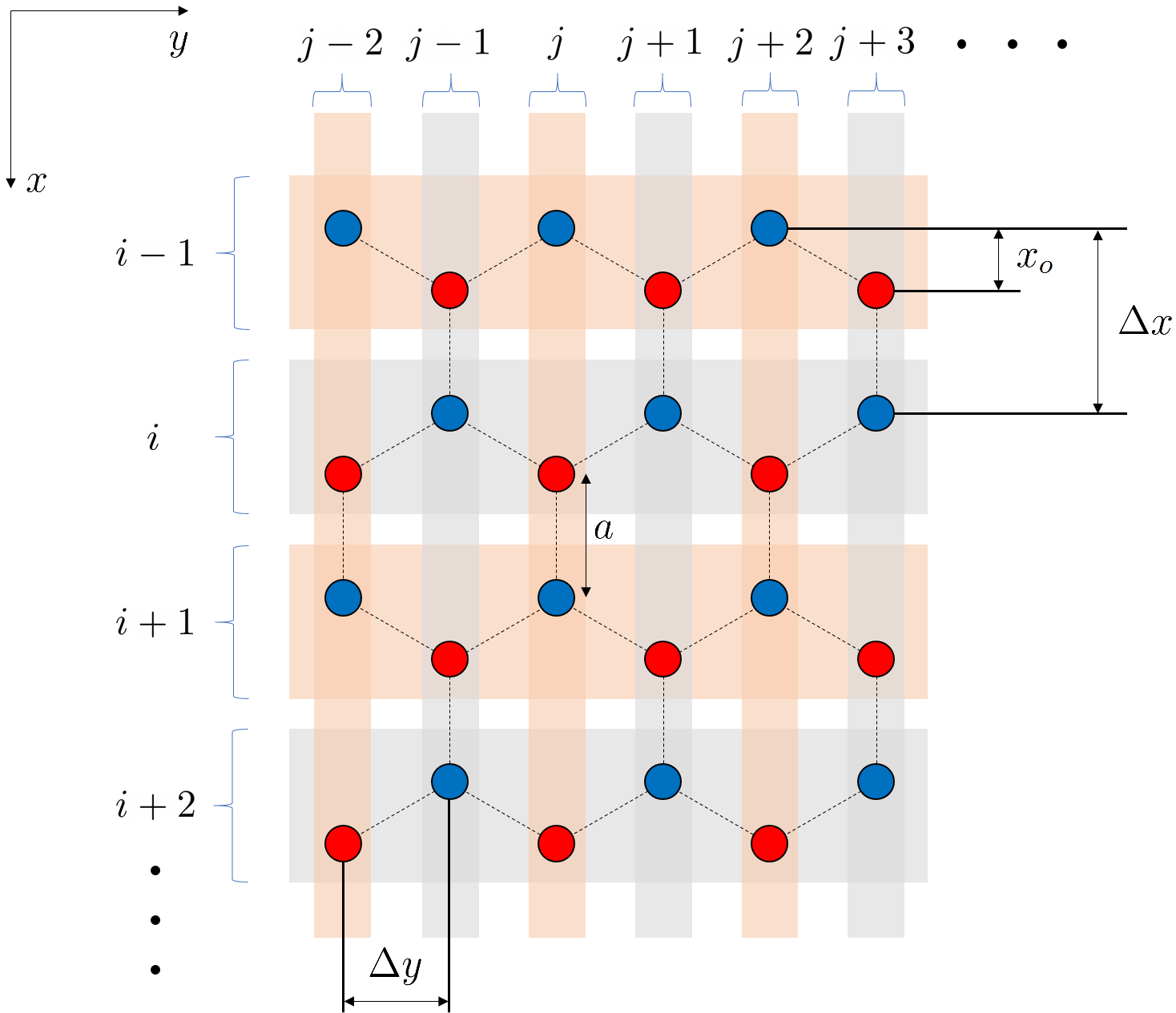}
    \caption{Sketch of the honeycomb lattice with parameter $a$. The lattice is bipartite and can be thus split into two sublattices: e-sites (blue), two nearest neighbours below and one above, and o-sites (red), two nearest neighbours above and one below. The distance between sites of the same kind corresponding to consecutive indexes are $\Delta x=3a/2$, $\Delta y=\sqrt{3}a/2$. The spacing between e- and o-sites with the same value of the index $i$ is $x_o=a/2$.}
    \label{fig:HoneycombLattice}
\end{figure}

A sketch of the lattice honeycomb lattice, with lattice constant $a$,
is shown in  Fig.~\ref{fig:HoneycombLattice}. Indices $i$ and $j$
are employed for rows and columns, respectively. Note that atoms at each
row are distributed in zigzag. There are two types of sites, e-sites
and o-sites.
For the former, the
three nearest neighbours are one above and two below; for the latter,
the three nearest neighbours are two above and one
below~\cite{ruiz-garcia_ripples_2015}.  Note that the 
two-dimensional domain $\Omega$ can have an arbitrary shape, not necessarily rectangular---for
the sake of simplicity, we assume that it has no holes. This means that all 
the nearest neighbours are present for the \textit{bulk} sites 
shown in  Fig.~\ref{fig:HoneycombLattice}---boundary sites will be taken into account by
introducing appropriate boundary conditions over the contour $\delta\Omega$ of the region $\Omega$,
as discussed later on.

The rotationally invariant Hamiltonian is the extension of the
spin-string one, as given by Eq.~\eqref{eq:energy-rot-inv}, to the
honeycomb lattice,
\begin{widetext}
\begin{align} \mathcal{H}(\hat{u},\hat{p},\hat{\sigma})&= \sum_{\abs{i-j}=\text{even}} \bigg[\frac{p_{ij}^2}{2m} +
\frac{k}{2}\left(u_{i-1,j}+u_{i,j-1}+u_{i,j+1}-3u_{ij}\right)^2  -h
\sigma_{ij}\left(u_{i-1,j}+u_{i,j-1}+u_{i,j+1}-3u_{ij}\right) \nonumber \\
 & \qquad \qquad \qquad + \frac{J}{2} \sigma_{ij} \left(\sigma_{i-1,j}+\sigma_{i,j-1}+
\sigma_{i,j+1}\right) \bigg]
\nonumber\\
  &+\sum_{\abs{i-j}=\text{odd}}
\bigg[ \frac{p_{ij}^2}{2m} + \frac{k}{2}\left(u_{i+1,j}+u_{i,j-1}+u_{i,j+1}-3u_{ij}\right)^2 -h
\sigma_{ij}\left(u_{i+1,j}+u_{i,j-1}+u_{i,j+1}-3u_{ij}\right) \nonumber \\
 & \qquad \qquad \qquad + \frac{J}{2} \sigma_{ij} \left(\sigma_{i+1,j}+\sigma_{i,j-1}+
\sigma_{i,j+1}\right) \bigg].
\end{align}
\end{widetext}
For the sake of clarity, the Hamiltonian is split into two sums because nearest neighbours are  different for e-sites and o-sites. Within each sum on the rhs, the first term stands for the kinetic energy, the second one corresponds to the elastic contribution to the Hamiltonian, the third one describes the interaction between the transverse displacements and the pseudospins, and the fourth one provides the antiferromagnetic coupling among the pseudospins.\footnote{The coupling constant is $J$, the factor $1/2$ is needed because each
interaction is counted twice.}

Again, we are interested in the equilibrium profiles of the membrane. Therefore, we integrate the canonical distribution over the momenta and the pseudospin variables, firstly introducing the continuum limit---see details in Appendix~\ref{app:2d-ContinuumLimit}. Now, the curvature is the extension of the one-dimensional case, i.e., the Laplacian of the displacements $\nabla^2 u(x,y)$. Then, the sum over $(i,j)$ goes to an integral over $(x,y)$, which leads to
\begin{equation}
  P_{\eq}[u]\propto e^{-\beta\calF[u]}, \quad \calF[u]=n \int_{\Omega}
  dx \, dy \, f(\chi),
\end{equation}
where  $n=(\Delta x \Delta y)^{-1}$ is the number density 
and $f$ is the free energy density per particle, given by
\begin{subequations}\label{eq:free-energy-density_2D}
\begin{align}
  f(\chi)&=\frac{k_{0}\chi^2}{2}
  -k_{B}T\ln\zeta_{\intern}^{\twodim}(\beta h_{0}\chi,\beta J),
\end{align}
\end{subequations}
$\zeta_{\intern}^{\twodim}(\beta h_{0}\chi,\beta J)$ stands for the partition
function for the antiferromagnetic Ising chain
with coupling $J$ and external field $h_{0}\chi$ in the
two-dimensional case. Unfortunately, at variance with the
one-dimensional case, there is not an analytical expression for
$\zeta_{\intern}^{\twodim}$ when $J\ne 0$.

The equilibrium profile minimises the free energy, so it is determined by the condition 
$\delta F[u] = 0$. Since the free energy density only depends on the curvature $\chi$, it is convenient to define the field
\begin{equation}\label{eq:Phi-def}
    \Phi(x,y)\equiv \pdv{f}{\chi}=k_0\chi - h_0\pdv{\ln\zeta_{\intern}^{\twodim}(\beta h_{0}\chi,\beta J)}{(\beta h_0 \chi)}.
\end{equation}
Making use of this definition, the variation of the free energy can be written as
\begin{align}
    \delta F[u] =&
   \int_\Omega d{x}\, d{y} \,
   \Phi \,\nabla^2 (\delta u) \nonumber \\
   =& \int_\Omega d{x}\, d{y} \,\nabla\!\cdot\!\corch{ \Phi \, \nabla(\delta u)-\nabla \Phi \delta u}
   + \! \int_\Omega d{x}\, d{y} \, \nabla^2\Phi \delta u \nonumber \\
   =& \int_{\delta\Omega} ds \, \hat{\bm{n}}\!\cdot\!\pr{ \Phi \, \nabla(\delta u)-\nabla \Phi \,\delta u}
   + \int_\Omega d{x}\, d{y} \, \nabla^2\Phi \,\delta u.
   \label{eq:deltaF-2d}
\end{align}
In the contour integral along the boundary $\delta\Omega$, $ds$ is the length element and $\hat{\bm{n}}$ is the outward pointing unit normal. On the one hand, the surface integral in the second term on the rhs of Eq.~\eqref{eq:deltaF-2d} provides us with the Euler-Lagrange equation,
\begin{equation}\label{eq:LaplaceEquation}
    \nabla^2 \Phi=0,
\end{equation}
i.e. the Laplace equation for the field $\Phi(x,y)$. On the other hand, the contour integral in the first term gives us the boundary conditions: since each term must vanish separately, we have
\begin{equation}\label{eq:bc-2d-1st}
    \Phi \, \pdv{}{n}\delta u=0, \quad \delta u\, \pdv{}{n}\Phi =0,
\end{equation}
where $\partial/\partial n\equiv \hat{\bm{n}}\cdot\nabla$ is the normal derivative. For the sake of concreteness, and consistently with our analysis of the one-dimensional spin-string model, we consider supported boundary conditions: the value of the transversal displacements vanish at the boundaries,
\begin{equation}\label{eq:bc-2d-2nd}
    u(x,y)=0, \quad (x,y)\in\delta\Omega,
\end{equation}
but the value of the normal derivative $\pdv{}{n}\delta u=0$ at the boundary is free. Therefore, the coefficient of $\pdv{}{n}\delta u$ in Eq.~\eqref{eq:bc-2d-1st} has to vanish,
\begin{equation}\label{eq:bc-2d-3rd}
    \Phi(x,y)=0, \quad (x,y)\in\delta\Omega.
\end{equation}

The solution of the Laplace equation \eqref{eq:LaplaceEquation} with the homogeneous Dirichlet boundary condition \eqref{eq:bc-2d-3rd} is identically zero:
\begin{align}\label{eq:Euler-Lagrange-2d}
    k_0\chi_\eq- h_0\pdv{\ln\zeta_{\intern}^{\twodim}(\beta h_{0}\chi_\eq,\beta J)}{(\beta h_0 \chi_\eq)}=0,
\end{align}
where we have made  use of the definition of $\Phi$. Although we do not have an explicit expression for $\zeta_{\intern}^{\twodim}$, Eq.~\eqref{eq:Euler-Lagrange-2d} neatly tells us that the equilibrium curvature is constant, independent of $(x,y)$ over all the two-dimensional domain $\Omega$. Interestingly, the nondimensional variables introduced in the one-dimensional spin-string model, as given by Eq.~\eqref{eq:dimensionless-variables}, also work in the two-dimensional case---only now the dimensionless number density is $n^*=n\ell_0^2$. The dimensionless free energy and free energy density are
\begin{subequations}
\begin{align}
  \calF^{*}[u^{*}](J^{*},T^{*})&=n^* \int_{\Omega}  dx^{*}\, dy^{*}\,
  f^{*}(\chi^{*};J^{*},T^{*}), \\
 f^{*}(\chi^{*};J^{*},T^{*})&\equiv \frac{1}{2}{\chi^{*}}^{2}-T^{*}\ln
                              \zeta_{\intern}^{\twodim}(\beta^* \chi^{*},\beta^* \! J^{*}),
                              \label{eq:non-dim-free-energ-dens-2d}
\end{align}
\end{subequations}
where we have taken into account that $\beta h_0\chi=\beta^*\chi^*$, $\beta J=\beta^* J^*$. In what follows, we drop again the asterisks in the dimensionless variables to simplify our notation. Equation~\eqref{eq:Euler-Lagrange-2d} is written in dimensionless variables as 
\begin{equation}\label{eq:Euler-Lagrange-2d-nondim}
    \chi_\eq=\mu(\beta\chi_\eq,\beta J), \quad \mu(\beta\chi_\eq,\beta J)\equiv\pdv{\ln\zeta_{\intern}^{\twodim}(\beta\chi_\eq,\beta J)}{(\beta \chi_\eq)}.
\end{equation}
Note that $\mu$ is the local magnetisation of the two-dimensional pseudospin lattice. The equilibrium free energy $\calF_{\eq}$ is given by
\begin{equation}\label{eq:free-energy-equil-2d}
  \calF_{\eq}(J,T)\equiv \calF[u_{\eq}](J,T)=N
  f(\chi_{\eq};J,T),
\end{equation}
which is extensive, as expected on a physical basis.

\subsection{Buckled states}\label{eq:buckled-states-2d}

An important novel feature of our model is its rotational invariance. Indeed, as was the case of the one-dimensional spin-string model, the two-dimensional spin-membrane model analysed here is invariant under rotations. Let us consider a certain equilibrium profile $u(x,y)$, which thus solves the Euler-Lagrange equation~\eqref{eq:Euler-Lagrange-2d-nondim}. Note that $\tilde{u}(x,y)=u(x,y)+Ax+By+C$,
is another possible equilibrium profile with the same free energy, which only depends on the curvature $\nabla^2 u(x,y)$. 
Similarly to the one-dimensional case, the family of transformations defined by $Ax+By+C$ contains any small two-dimensional rotation.

To univocally determine the equilibrium profile for the membrane, we have to bring to bear the remainder of the supported boundary conditions, i.e. Eq.~\eqref{eq:bc-2d-2nd}. Therefore, we have to solve Poisson's equation, with the uniform curvature playing the role of the density and  homogeneous boundary conditions:
\begin{equation}\label{eq:Poisson-supported-bc}
    \nabla^2 u_\eq(x,y)=\chi_\eq,\; (x,y)\in\Omega; \;\; u(x,y)=0, \; (x,y)\in\delta\Omega.
\end{equation}
The solution for the profile depends on the geometry of the problem. The simplest situation appears for a circular membrane of radius $R$: therein, the equilibrium profile only depends on the distance $r$ to the centre of the membrane and straightforward integration of the Poisson equation gives
\begin{equation}\label{eq:eq-profile-circle}
    u_\eq(r)=\frac{\chi_\eq}{4}(r^2-R^2).
\end{equation}
For a rectangle of sides $L_x$ and $L_y$, the solution of the Poisson equation involves an expansion in eigenfunctions---for details, see Appendix~\ref{app:profile-rectangular-domain}. 

Below, we analytically investigate the behaviour of the two-dimensional spin-membrane model in some limits of interest, for which the expression for $\zeta_{\intern}^{\twodim}$ can be worked out. Specifically, two situations are considered: (i) the case $(J=0,T\ne 0)$, in which the antiferromagnetic coupling among the pseudospins disappears, and (ii) the low-temperature limit $(J\ne 0,T=0)$. In the former, we show that there appears a second-order phase transition at $T=1$ (in dimensionless variables), below which the zero curvature membrane is unstable and stable buckled profiles bifurcate. In the latter, we show that the zero curvature membrane is always locally stable but membrane buckled states (one locally stable, another unstable) are present below a certain value $J_M^{(0)}$ of the antiferromagnetic coupling. Moreover, metastability is present, signaling that the transition from zero curvature to buckled states is now first-order. 

\subsubsection{No antiferromagnetic coupling}\label{sec:J-zero-2d} 

Let us first consider the case $J=0$. The one-particle partition $\zeta_{\intern}^{\twodim}(J=0,T)$ corresponds to that of a non-interacting Ising system in an external field $\chi$, i.e.
\begin{align}
    f(\chi;J=0,T)=\frac{1}{2}\chi^2-T \ln \corch{2\cosh\pr{\frac{\chi}{T}}}.
\end{align}
and the Euler-Lagrange for the curvature reduces to
\begin{equation}\label{eq:EL-2d-Jzero}
    \chi_{\eq}(J=0,T) =\tanh\corch{\frac{\chi_{\eq}(J=0,T)}{T}} .
\end{equation}
This coincides with the particularisation of Eq.~\eqref{eq:profile-eq-nondim} for $J=0$---this is logical, without antiferromagnetic interaction the transition is of mean field type for any dimension. Therefore, the ZC phase is the only one for $T>1$ and is stable, whereas it becomes unstable for $T<1$. At $T=1$, a stable buckled phase B continuously bifurcates from the ZC phase, namely
\begin{equation}
    \chi_\eq \sim \sqrt{3(1-T)}, \quad  \text{for} \  T\to 1^-.
\end{equation}
This phase B is the stable one for $T<1$ and continues to exist up to $T=0$. In fact,  Eq.~\eqref{eq:EL-2d-Jzero} predicts that
\begin{equation}\label{eq:chi-J0-2d-lowT}
   \lim_{T\to 0^+}\chi_\eq(J=0,T)= 1. 
\end{equation}

\subsubsection{Low-temperature limit}\label{sec:low-temp-2d}

Now we turn to the low-temperature limit $T\to 0^+$, with $J\ne 0$. In this regime, similarly to the situation in the one-dimensional spin-string system, we have
\begin{equation}\label{eq:egs-def-2d}
   \lim_{T\to 0^+}-T\ln\zeta_{\intern}^{\twodim}=e_{\gs}^{\twodim}
\end{equation}
where
\begin{equation}\label{eq:egs-expression-2d}
    e_{\gs}^{\twodim}=-\pr{\left|\chi\right|-3J}H\pr{\left|\chi\right|-3J}-\frac{3J}{2}.
\end{equation}
is the energy per site in the ground state of the two-dimensional spin-membrane. Again, Eq.~\eqref{eq:egs-expression-2d}  for $e_{\gs}^{\twodim}$ has a clear physical interpretation. On the one hand, the ground state energy corresponds to antiferromagnetic ordering for small $\chi$, $e_{\gs}^{\twodim}=-3J/2$---in the honeycomb lattice, each pseudospin has three nearest neighbours. On the other hand, the ground state corresponds to all pseudospins aligned with the curvature for large $\chi$, $e_{\gs}^{\twodim}=-\chi+3J/2$. The crossing between these expressions for $e_{\gs}^{\twodim}$ takes place at $\chi=3J$, which leads to Eq.~\eqref{eq:egs-expression-2d}---see details in Appendix~\ref{app:GS-2d}. 

The free energy density is now
\begin{equation}\label{eq:free-energ-dens-low-T-2d}
    f(\chi;J,T)\sim\frac{\chi^2}{2}
    -\pr{\left|\chi\right|-3J}H\pr{\left|\chi\right|-3J}-\frac{3J}{2}, \quad T\ll 1,
\end{equation}
and the equilibrium curvature is fixed by
\begin{align}\label{eq:EL-low-T-2d}
    \abs{\chi_{\eq}}=H(\abs{\chi_{\eq}}-3J), \quad
  T\ll 1.
\end{align}
We recall that, without loss of generality, we are restricting ourselves to positive curvatures in the paper, thus $\abs{\chi_{\eq}}\to\chi_{\eq}$ henceforth. The analysis of the low-temperature limit follows along the same lines as in the one-dimensional case, so we present the results in a concise way. The ZC phase is always a solution of Eq.~\eqref{eq:EL-low-T-2d}, and two buckled solutions emerge for $3J\leq 1$, i.e. $J\leq J_{M}^{(0)}= 1/3$. The two buckled phases have curvatures $\chi_{\eq}=3J$ and $\chi_{\eq}=1$. For $J>J_M^{(0)}$, the antiferromagnetic interaction is so strong that prevents the emergence of buckled solutions. 

To identify the phases, we evaluate the free energy density over them, the ZC phase verifies  $f_{\zerocurv}(J,T)\sim -3J/2$, for $T\ll 1$.  Therefore
\begin{align}
  \Delta f(\chi;J,T) &=
  f(\chi;J,T)-f_{\zerocurv}(J,T) \nonumber \\ 
  &\sim
  \frac{\chi^2}{2}
  -\pr{\left|\chi\right|-3J}H\pr{\left|\chi\right|-3J},
  \quad T\ll 1.
\end{align}
and
\begin{subequations}
\begin{align}
    \Delta f(\chi_{\eq}=3J; J,T)&\sim \frac{9}{2}J^{2}, \\
    \Delta f(\chi_{\eq}=1;J,T)&\sim -\frac{1}{2}+3J.
\end{align}
\end{subequations}
The phase with $\chi_{\eq}=3J$ always has $\Delta f>0$. The phase with $\chi_{\eq}=1$ has $\Delta f<0$ ($>0$) for $0\leq J<J_{t}^{(0)}=1/6$ $\left(J_{t}^{(0)}<J\leq J_{M}^{(0)}\right)$.  Then, the phase with $\chi_{\eq}=3J$ should correspond to the (unstable) B- phase, whereas the phase with $\chi_{\eq}=1$ should correspond to the (at least locally stable) B+ phase. The buckled phases disappear discontinuously at $J=J_M^{(0)}$, which means that the transition is first-order.

The above identification of the phases is physically sensible, since in the limit $J\to 0$ the B+ phase for $T\ll 1$ is expected to converge to the $T\to 0^+$ limit of the B phase found for $J=0$---and, in fact, the curvature of the latter has the right behaviour, as expressed by Eq.~\eqref{eq:chi-J0-2d-lowT}. The profile of the B+ phase for a circular membrane of unit area is shown in Fig.~\ref{fig:buckled-lowT-2d}; the anagolous profile for a rectangular membrane can be found in Appendix~\ref{app:profile-rectangular-domain}.
\begin{figure}
    \centering
    \includegraphics[width=3.25in]{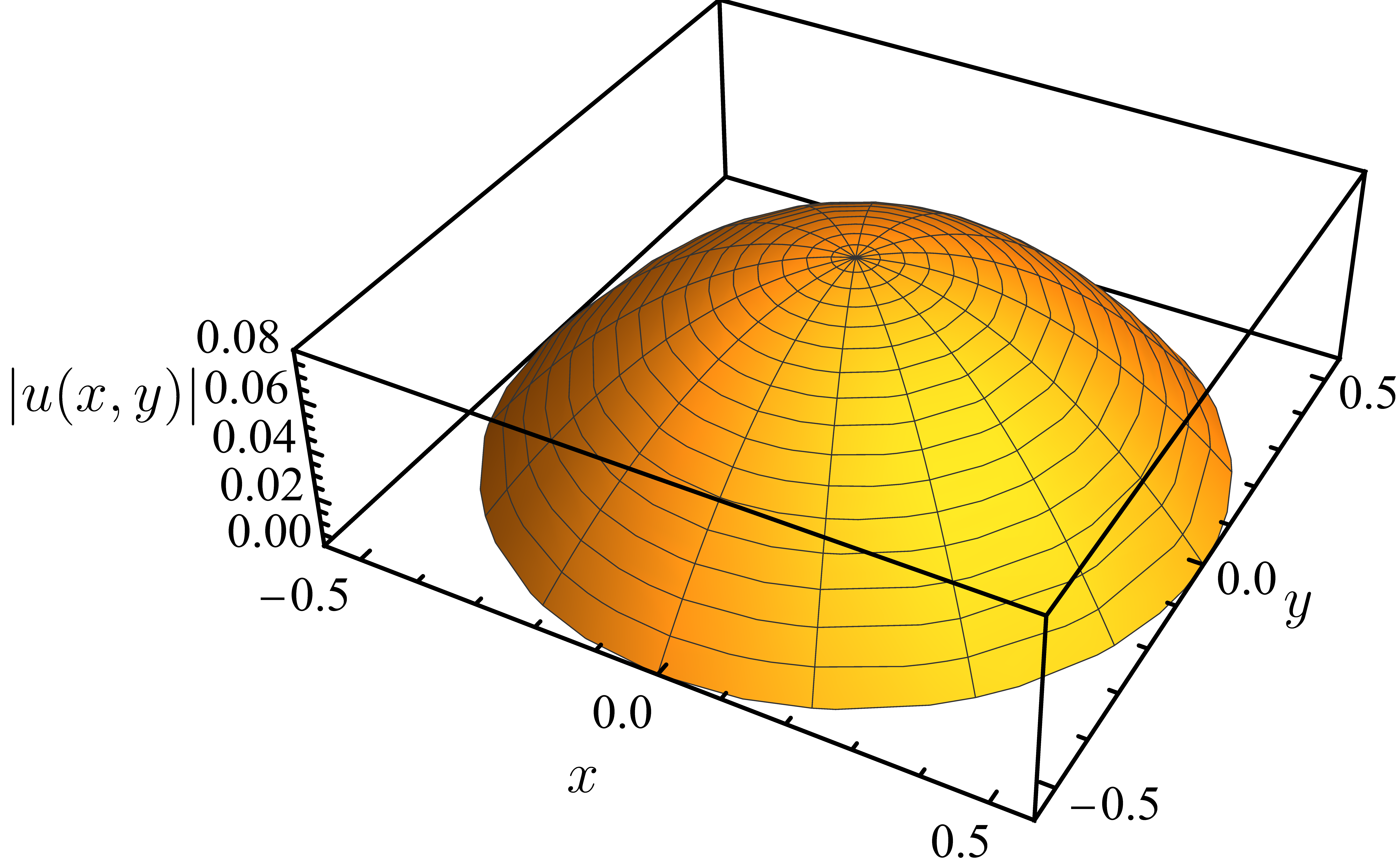}
    \caption{Low-temperature stable buckled configuration for the spin-membrane model.  The plotted surface corresponds to the absolute value of Eq.~\eqref{eq:eq-profile-circle}, for a unit area circle. The curvature equals unity for all $J\leq J_M^{(0)}=1/3$, corresponding to the most stable state for $0\leq J<J_t^{(0)}=1/6$ and to a metastable state for $J_t^{(0)}<J\leq J_M^{(0)}$.}
    \label{fig:buckled-lowT-2d}
\end{figure}

\section{Conclusions}\label{sec:conclusions}

In this paper, we have introduced novel, rotationally invariant, spin-string and spin-membrane models. The main role is played by the curvature associated with the transversal displacements, which not only rules the elastic energy but also the coupling with internal degrees of freedom.  These models give an appealing physical picture of the buckling transition while keeping consistency with both thermodynamics and classical theory of elasticity, which is important from a theoretical perspective. Moreover, the equilibrium profiles are very simple, with a position-independent curvature, both for the one-dimensional and the two-dimensional lattices.  

The phase diagram in the $(J,T)$ plane---$J$ being the antiferromagnetic coupling among the pseudospins and $T$ the system temperature---of the one-dimensional rotationally invariant spin-string model has been characterised analytically  in great detail. It is complex and qualitatively similar to that found in Ref.~\cite{ruiz-garcia_bifurcation_2017} and, in particular, buckled phases emerge for low enough values of the temperature and the antiferromagnetic coupling. In absence of the latter, there appears a second-order phase transition at a certain temperature, below which the zero curvature phase becomes unstable and a buckled B phase is the stable one. In the buckled phase, the pseudospins magnetisation equals the non-zero curvature and, thus, they predominantly align with the sign of the curvature. For low temperatures, the ferromagnetic ordering of the pseudospins is frustrated by the antiferromagnetic coupling, which destroys the buckled states for high enough coupling constants $J$, $J>J_M^{(0)}$. For $J\leq J_M^{(0)}$, there appears coexistence between the ZC phase and a stable B+ buckled phase, phase B+ is the most stable one and phase ZC is metastable for small $J$, $0\leq J<J_t^{(0)}$, with the situation being reversed for $J_t^{(0)}<J\leq J_M^{(0)}$. This qualitative map of phases is the key ingredient that makes this kind of models qualitatively explain the transition from a rippled to a buckled state when the temperature is increased~\cite{ruiz-garcia_stm-driven_2016,schoelz_graphene_2015}.

Despite our qualitative explaining of the experimentally observed rippled to buckled transition, the model presented here does have some limitations. The main one stems from our starting from a mesoscopic description. Our focus is thus the characterisation of the kind of physical interactions that are able to cause the observed phenomenology, and not a quantitative account of the mechanical properties of the graphene. In order to achieve this latter goal, one would have to start from a microscopic description and go to a mesoscopic model by taking an adequate physical limit, which is outside the scope of the present work. Nevertheless, it is precisely the mesoscopic nature of our modelling that allows us to obtain analytical results and a complete characterisation of the different phases.

To be concrete, and motivated by the geometry of graphene, we have studied the spin-membrane model in a honeycomb lattice. The analytical expression of the partition function for the pseudospins is not known in the two-dimensional case and thus the phase diagram cannot be characterised as completely as in the one-dimensional case. Nevertheless, our analysis of limiting cases allows us to state that the same qualitative picture of phases applies for the two-dimensional spin-membrane model. Still, the two-dimensional values of $J_t^{(0)}$ and $J_M^{(0)}$ are smaller---in dimensionless variables---than the one-dimensional ones. This can be understood as stemming from the antiferromagnetic interaction being stronger, since the number of nearest neighbours is larger for $d=2$ than for $d=1$. It should be stressed that the main conclusions of our study of the two-dimensional case hold for a quite general lattice, as long as it does not contain triangular loops---such as a rectangular one.

Previous spin-string and spin-membrane models~\cite{ruiz-garcia_stm-driven_2016,ruiz-garcia_bifurcation_2017,ruiz-garcia_ripples_2015} had several weaknesses. First, it was necessary to assume the parameters in the system Hamiltonian to have certain scalings with system size to get consistency with thermodynamics, specifically extensive free energy and transition temperature independent of system size. Second, even assuming size-dependent microscopic parameters, the models were not consistent with classical elasticity theory: e.g. a flat but rotated string or membrane was not an acceptable equilibrium profile.

The models developed in this work mend the aforementioned weaknesses. The microscopic parameters $k$ and $h$ do not depend on system size and determine natural units for length and temperature, which we have used to introduce dimensionless variables. From them, we have naturally obtained a system-size-independent transition temperature and an extensive free energy. Moreover, the Euler-Lagrange equation is consistent with the classical theory of elasticity, with the term stemming from the internal interactions involving a biLaplacian instead of a Laplacian. These two properties hold for both the one-dimensional and two-dimensional lattices. 

Note that there are no external fields in our Hamiltonian, and therefore buckling here is a purely thermal phenomenon---there is no way of making the system buckle as a consequence of  mechanical actions. For instance, it would be interesting to consider a prescribed force applied to one extremity of the spin-string chain (or to the border of the spin-membrane). From the point of view of statistical mechanics, two different statistical ensembles emerge (Gibbs-like, controlled force, and Helmholtz-like, controlled length or area), which are relevant both from a theoretical standpoint and for practical applications.  The continuum limit considered here does not need---as already discussed, and at variance with previous works---the system to be large. Therefore, the problem of ensemble equivalence, which has been analysed in different contexts~\cite{barre_large_2005,winkler_equivalence_2010,manca_equivalence_2014,prados_sawtooth_2013,bonilla_theory_2015,benedito_isotensional_2018}, is also worth studying here---both ensembles are expected to be equivalent only in the thermodynamic (large system size) limit.

Also, perspectives for future work are opened. In the context of elasticity of low-dimensional systems, our work may trigger new approaches to the analysis of buckled phases in more realistic models. Also, it would be interesting to consider variants of the two-dimensional model that allow for a more detailed analytical treatment. In this regard, effective antiferromagnetic interactions for which the pseudospins partition function is exactly known~\cite{fisher_lattice_1960} deserve to be studied. Moreover, the framework developed here may be useful to look into the elastic behaviour of systems in which elastic modes are coupled to other, internal, degrees of freedom---such as biomolecules~\cite{peyrard_biophysics:_2006,prados_spin-oscillator_2012,benedito_isotensional_2018,cannizzo_thermal_2022}. A different but also interesting prospect is the investigation of dynamics, since the present study only concerns equilibrium configurations. On a physical basis, one expects the dynamics of the internal degrees of freedom to be much faster than that of the elastic modes. This separation of time scales should make it possible---e.g. making use of a Chapman-Enskog expansion~\cite{bonilla_chapman-enskog_2000,baldovin_derivation_2019}---to derive an effective stochastic dynamics, either at the Langevin or Fokker-Planck level of description,  for the elastic modes. 

\appendix

\section{Bifurcation theory details}\label{sec:bifurc-theory}
In this Appendix, the different regions of the system phase diagram are derived in detail using the expansion of the free energy, through bifurcation theory.  

Equation~\eqref{eq:free-energy-expansion} allows us to analyse the
emergence of buckled ($\chi\neq 0$) profiles from the zero-curvature
solution, using the Landau theory of phase transitions---with the
curvature playing the role of the order parameter.  Within this
approximation, the transcendental Euler-Lagrange
equation~\eqref{eq:profile-eq-nondim} for the curvature becomes algebraic: $\left.\pdv{\Delta f(\chi;J,T)}{\chi}\right|_{\eq}=0$ implies
\begin{align}\label{eq:profile-eq-nondim-expansion}
0=&  
 f_{2}(J,T)\chi_{\eq}+
  \frac{f_{4}(J,T)}{3!}\chi_{\eq}^{3} 
 + \frac{f_{6}(J,T)}{5!}\chi_{\eq}^{5} .
\end{align}
This expansion is truncated at different orders,
the sign of the last retained coefficient $f_{n}$ must be always
positive---otherwise, the corresponding equilibrium distribution
$P_{\eq}$ would be non-normalisable.

\subsubsection{ \texorpdfstring{$f_{4}>0$}{f40}: Second-order phase transition}
\label{sec:2nd-order-phase-trans}

When $f_4>0$, the sign of $f_2$ determines the possible phases of the system.  Close to any point $(J_{b},T_{b})$ over the bifurcation curve~\eqref{eq:bifurc-curve}, $f_{2}$ is small: we explicitly indicate this by introducing a parameter $0<\eps\ll 1$ such that $f_{2}=\eps\varphi_{2}$, with $\varphi_{2}=O(1)$. This tells us that the separation of the point $(J,T)$ we are considering from the bifurcation curve is of order
$\eps$: either $T-T_{b}=O(\eps)$ or $J-J_{b}=O(\eps)$. We can take advantage of these results to analyse our model in the ferromagnetic case $J<0$. Therein, no tricritical point stemming from the competition of the interactions emerge, because $f_4(J,T) > 0$ for any $J\leq 0$. Therefore, the bifurcation curve inside the second quadrant $(J<0,T>0)$ defines a typical second-order phase transition line. In order to clarify how the regions are extended to the ferromagnetic situation, a small portion of the second quadrant is included in Fig.~\ref{kappavstheta3}.

On the one hand, if $\varphi_2>0$, the free energy density has only
one minimum at $\chi_{\eq}^{\zerocurv}=0$. On the other hand, if
$\varphi_{2}<0$, $\chi_{\eq}^{\zerocurv}=0$ turns out to be a maximum and
there appear two symmetric minima at
\begin{equation}\label{eq:chieq-2nd-order}
  \chi_{\eq}^{\text{B}}=\pm\varepsilon^{1/2}\sqrt{-6\varphi_{2}/f_{4,b}},
\end{equation}
where $f_{4,b}\equiv f_{4}(J_{b},T_{b})$ is the value of the
coefficient $f_{4}$ over the bifurcation curve,
\begin{equation}
  f_{4,b}=\frac{3T_{b}^{2}-1}{T_{b}^2}.
\end{equation}
The superindex B in Eq.~\eqref{eq:chieq-2nd-order} indicates that this
solution corresponds to a buckled phase, with non-zero curvature.
Inside region II of Fig.~\ref{kappavstheta3}, the stable profile for
the string is thus buckled.

Moreover, the term we have neglected in
Eq.~\eqref{eq:profile-eq-nondim-expansion}, $f_{6}\chi_{\eq}^{5}/6!$
is in fact negligible: it is of order $\eps^{5/2}$, whereas the first
two terms on the rhs are of order $\eps^{3/2}$---this shows the
consistency of the approximations introduced.

\subsubsection{\texorpdfstring{$f_{4}<0$}{f40} and \texorpdfstring{$f_{6}>0$}{f60}: Tricritical point and first-order phase transition}\label{sec:1st-order-phase-trans}

The coefficient $f_{4,b}$ over the bifurcation curve vanishes at the
tricritical point $C\equiv(J_{c},T_{c})$, given by
\begin{equation}
    T_c=\frac{1}{\sqrt{3}}, \quad J_c=J_b(T_c)=\frac{\ln{3}}{4\sqrt{{3}}}.
\end{equation}
Close to the tricritical point, the approximation given by
Eq.~\eqref{eq:chieq-2nd-order} thus breaks down. In fact, below the
tricritical point---over the leftmost dotted line in
Fig.~\ref{kappavstheta3}---we have that $f_{4,b}<0$ and  the term
involving $f_{6}$ in the expansion of the free energy density must be
retained.

Close to the tricritical point, a different scaling for the solutions
of Eq.~\eqref{eq:profile-eq-nondim-expansion} is needed. We still
write $f_{2}=\eps\varphi_{2}$: the terms involving $f_{2}$ and $f_{6}$
are of the same order if $\chi_{\eq}=O(\eps^{1/4})$. Then, the terms
involving $f_{2}$ and $f_{4,b}$ are of the same order if
$f_{4,b}=O(\eps^{1/2})$---this tells us the order of the distance of the
point $(J_{b},T_{b})$ over the bifurcation curve to the tricritical point, i.e. either
$T_{b}-T_{c}=O(\eps^{1/2})$ or $J_{b}-J_{c}=O(\eps^{1/2})$. Then, we write
$f_{4,b}=\eps^{1/2}\varphi_{4}$, $f_{6,c}=f_{6}(J_{c},T_{c})=36$, and the
dominant balance for the curvature equation in this region is
\begin{equation}\label{eq:curv-cond-tricrit}
  \varphi_{2}\Xi+\frac{1}{6}\varphi_{4}\Xi^{3}+
  \frac{1}{120}f_{6,c}\Xi^{5}=0,  \qquad \chi_{\eq}\equiv\eps^{1/4}\Xi.
\end{equation}
We find the zero-curvature solution $\chi_{\eq}^{\zerocurv}=0$, and two
buckled solutions
\begin{subequations}
 \begin{align}
  \chi_{\eq}^{\BuckledStable}=&\pm\eps^{1/4}\sqrt{\frac{-5\varphi_4+
                    \sqrt{25\varphi_4^2-30\varphi_2f_{6,c}}}{f_{6,c}/2}}, \label{eq:chi-B+}
  \\ \chi_{\eq}^{\BuckledUnstable}=&\pm\eps^{1/4}\sqrt{\frac{-5\varphi_4- \sqrt{25\varphi_4^2-30\varphi_2f_{6,c}}}{f_{6,c}/2}}.
\end{align}   
\end{subequations}
The zero-curvature phase corresponds to a local minimum of the free
energy as long as $f_{2}>0$, i.e. outside the bifurcation
curve~\eqref{eq:bifurc-curve}---it becomes unstable inside it, as
already discussed. The curvatures $\chi_{\eq}^{\text{B}\pm}$ correspond to
two buckled phases B$\pm$, the domain of existence and stability of
which is discussed below.

The buckled phase B+ exists as long as $f_{4,b}<0$ (or $\varphi_{4}<0$),
i.e. below the tricritical point, provided that
$5f_{4,b}^{2}-6f_{2}f_{6,c}>0$ (or
$5\varphi_4^2-6\varphi_2f_{6,c}>0$). The line
\begin{equation}\label{eq:end-metastab-v1}
  5f_{4,b}^{2}-6f_{2}f_{6,c}=0
\end{equation}
marks the end of the existence of this phase (rightmost dotted curve in Fig.~\ref{kappavstheta3}). Moreover, when traversing from right to left this line, the buckled phase pops up discontinuously with a finite curvature: specifically, from
$\chi_{\eq}^{\zerocurv}=0$ to $\chi_{\eq}^{\BuckledStable}=\pm \sqrt{-5f_{4,b}/18}\ne 0$.  It can be easily checked---by evaluating the
second derivative of the free energy density with respect to $\chi$---that this buckled phase is locally stable in its domain of existence.

The buckled phase B- also exists as long as $f_{4,b}<0$ (or
$\varphi_{4,b}<0$), i.e. below the tricritical point. However, not only
does it need $5f_{4}^{2}-6f_{2}f_{6,c}>0$ 
but also $f_{2}>0$. 
This phase B- only exists between the line~\eqref{eq:end-metastab-v1}  and the part of the bifurcation curve~\eqref{eq:bifurc-curve} below the tricritical point (dotted lines in Fig.~\ref{kappavstheta3}). Phase B- is always unstable: the second derivative of the free energy density for $\chi=\chi_{\eq}^{\BuckledUnstable}$ is negative.

The curve demarcating the limit of existence of the buckled phases,
Eq.~\eqref{eq:end-metastab-v1}, can be written in a more transparent
way. Let us consider, for any $T$ close to $T_{c}$, a point
$(J,T)$ close to the bifurcation curve $(J_{b}(T),T)$ by taking into account that
\begin{equation}\label{eq:f4b-f2-tricrit}
  f_{4,b}\sim 6\sqrt{3} (T-T_{c}), \quad f_{2}\sim 2\sqrt{3} (J-J_{b}(T))
\end{equation}
in the approximation we are employing. Therefore,
Eq.~\eqref{eq:end-metastab-v1} is equivalent to
\begin{equation}\label{eq:limit-metastab-curve}
  J=J_{M}(T)\equiv J_{b}(T)+\frac{5}{12}\sqrt{3} (T-T_{c})^{2}.
\end{equation}

The analysis above entails the emergence of the region III in the
plane $(J,T)$, also depicted in Fig.~\ref{kappavstheta3}. This region
extends over the zone of the plane $(J,T)$ below the tricritical point
$C$, with its left border at the bifurcation line~\eqref{eq:bifurc-curve} and its right border
at the line~\eqref{eq:limit-metastab-curve} (both dotted lines in Fig.~\ref{kappavstheta3}). Inside region III, the three phases \zerocurv, B+, and B- coexist. Phases B+ and \zerocurv\ correspond to local
minima of the free energy, whereas B- always corresponds to a local
maximum. The relative stability of phases B+ and \zerocurv\ is elucidated in
the following.

In region III, the phases \zerocurv\ and B+ are both locally stable: one of
them corresponds to the deepest minima, being the most favourable
thermodynamic state, whereas the other one corresponds to a metastable
state. This means that the
transition changes from second-order above the tricritical point to
first-order below it, since the curvature (order parameter) of the most stable phase changes abruptly below the tricitical point. The change of stability takes place at the first-order
transition line determined by the condition $\Delta \calF(\BuckledStable)=\Delta
\calF(\zerocurv)=0$.  Close to the tricritical point, this is equivalent to
\begin{equation}
  f_{2}+\frac{1}{12}f_{4,b}(\chi_{\eq}^{\BuckledStable})^{2}+
\frac{1}{360}f_{6,c}(\chi_{\eq}^{\BuckledStable})^{4}=0.
\end{equation}
Bringing to bear Eqs.~\eqref{eq:curv-cond-tricrit} and
\eqref{eq:chi-B+}, over the first-order line one has
$8f_{2}f_{6,c}=5f_{4,b}^{2}$ that translates into
\begin{equation}\label{eq:first-order-curve}
  J=J_{t}(T)\equiv J_{b}(T)+\frac{5}{16}\sqrt{3}(T-T_{c})^{2},
\end{equation}
making use of Eq.~\eqref{eq:f4b-f2-tricrit}. The buckled phase B+ is
the most stable one in region IIIa of Fig.~\ref{kappavstheta3},
i.e. between the branch of the bifurcation line below the tricritical
point (leftmost dotted line) and the first-order line
\eqref{eq:first-order-curve} (dashed line), with the \zerocurv\ phase
being metastable. The situation is just reversed in region IIIb between the
first-order line \eqref{eq:first-order-curve} and the curve
demarcating the limit of existence of  phases B$\pm$,
Eq.~\eqref{eq:limit-metastab-curve} (rightmost dotted line): therein, the
\zerocurv\ phase is most stable and B+ is metastable.

\section{Continuum limit in honeycomb lattice}\label{app:2d-ContinuumLimit}
We go to a continuum limit, by assuming that $u_{ij}$ varies
slowly with $(i,j)$. We introduce continuous spatial variables
$x$ and $y$ as depicted in Fig.~\ref{fig:HoneycombLattice}, 
\begin{subequations}
\begin{align} x=&i \Delta x, & y=&j \Delta y , &
\text{e-sites sublattice}, \\ x=&x_o + i \Delta x, & y=&j \Delta y, &
\text{o-sites sublattice}.
\end{align}
\end{subequations}
where we have defined $x_o=a/2$, $\Delta x=3a/2$, and $\Delta y=\sqrt{3}a/2$. All nearest neighbours of o-sites are e-sites, and vice versa---the honeycomb lattice is bipartite~\cite{newman_networks_2010}. Therefore, the term in parentheses of the elastic term for e-sites is, neglecting $O(a^3)$ terms,
\begin{align}
  u_{i+1,j}+&u_{i,j-1}+u_{i,j+1}-3u_{ij} \nonumber \\
            &=u(x+l,y)+u(x-l/2,y-a/2) \nonumber \\ 
            & \quad +u(x-l/2,y+a/2)-3u(x,y)\nonumber \\
  &\sim \left(\sqrt{3}\frac{a}{2}\right)^{2}\chi(x,y), \qquad \chi(x,y)\equiv \nabla^{2}u(x,y).
\end{align}
where, in analogy with the one-dimensional case, we have introduced the notation $\chi$ for the curvature of the membrane. The same result holds for o-sites, $u_{i-1,j}+u_{i,j-1}+u_{i,j+1}-3u_{ij}\sim \left(\sqrt{3}a/2\right)^{2}\chi(x,y)$.   Then, we get
\begin{subequations}
\begin{align}
    \frac{k}{2}(u_{i+ 1,j}+u_{i,j-1}+u_{i,j+1}-3u_{i,j})^2 &= \frac{k_0}{2}\chi^2, \\
    h(u_{i+1,j}+u_{i,j-1}+u_{i,j+1}-3u_{i,j})\sigma_{ij}&= h_0\chi\sigma_{ij},
\end{align}
\end{subequations}
in which we have defined $k_0=k(\sqrt{3}a/2)^4$ and $h_0=h(\sqrt{3}a/2)^2$.

\section{Buckled profile on a rectangular domain}\label{app:profile-rectangular-domain}

\begin{figure}
    \centering
    \includegraphics[width=3.25in]{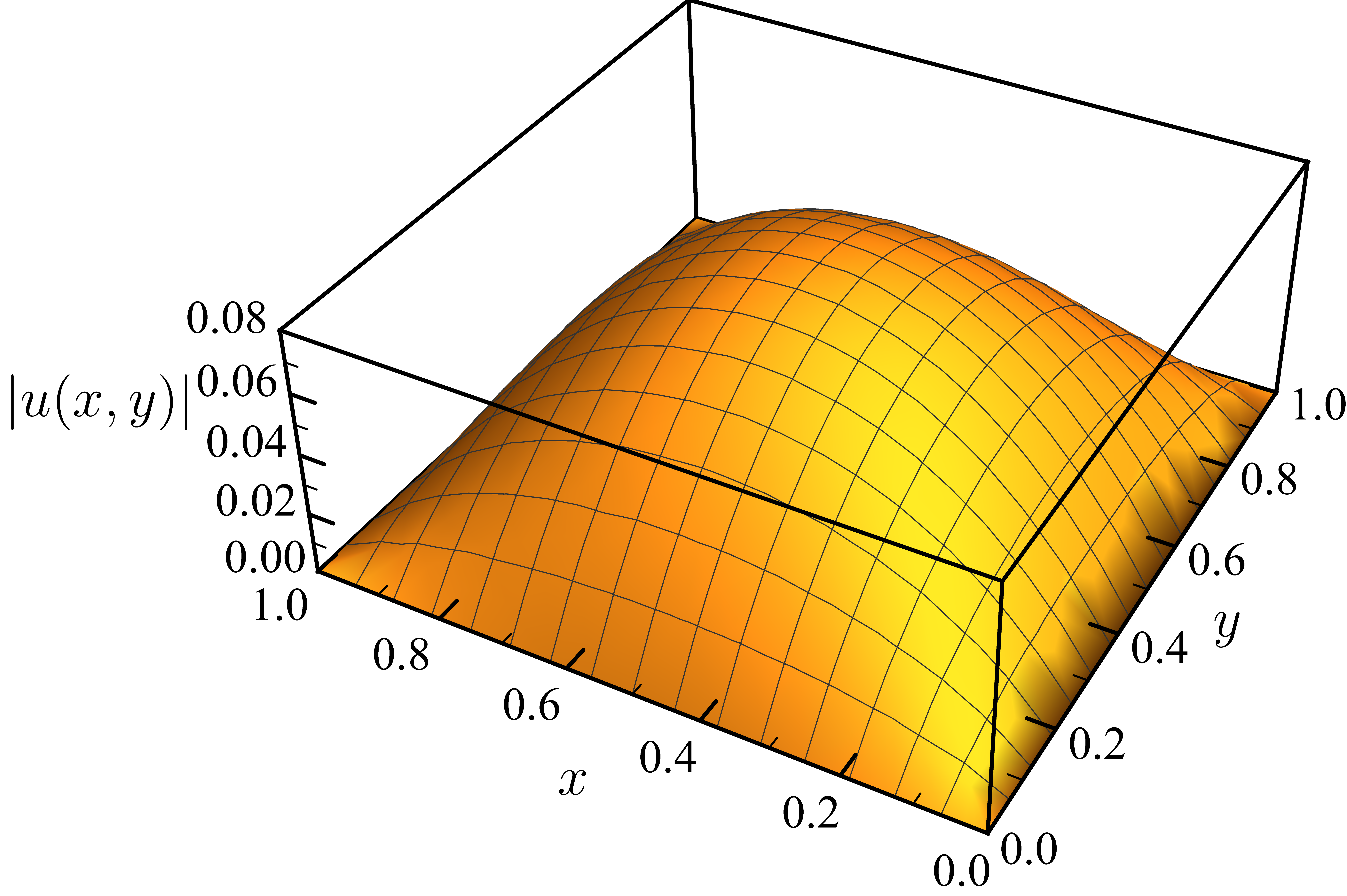}
    \caption{Absolute value of the stable buckled state on a square domain.  The graph corresponds to the low-temperature limit, in which $\chi_\eq=1$, for a unit area square with supported boundary conditions.}
    \label{fig:buckled-profile-rectangle}
\end{figure}

Here, we derive the analytical expression for the solution of Poisson's equation \eqref{eq:Poisson-supported-bc} on a rectangular domain of sides $L_x$ and $L_y$. We propose the solution
\begin{equation}
    u(x,y) = \chi_{\text{eq}} \left[ \frac{x(x-L_x)}{2} + w(x,y) + z(x,y)\right].
\end{equation}
The first term on the rhs verifies Poisson's equation and the homogeneous Dirichlet boundary conditions at the sides $x=0$ and $x=L_x$, but not at the sides $y=0$ and $y=L_y$. That is the reason why we introduce the functions $w(x,y)$ and $z(x,y)$; both satisfy Laplace's equation but with different boundary conditions, specifically
\begin{subequations}
\begin{align}
        w(0,y)&=w(L_x,y)=w(x,0)=0,  \\
        w(x,L_y)&=-x(x-L_x)/2, \\
        z(0,y)&=z(L_x,y)=z(x,L_y)=0, \\ 
        z(x,0)&=-x(x-L_x)/2,
\end{align}
\end{subequations}
so that $u(x,0)=u(x,L_y)=0$. Applying separation of variables and imposing the boundary conditions, we obtain
\begin{align}
    w(x,y) &= \frac{4L_x^2}{\pi^3}\sum_{n \scriptsize{\text{ odd}}} \frac{\sinh(\frac{n \pi y}{L_x})}{\sinh(\frac{n \pi L_y}{L_x})}\frac{\sin\pr{\frac{n\pi x}{L_x}}}{n^3}, \\
    z(x,y) &= \frac{4L_x^2}{\pi^3}\sum_{n \scriptsize{\text{ odd}}} \frac{\sinh(\frac{n \pi (L_y-y)}{L_x})}{\sinh(\frac{n \pi L_y}{L_x})}\frac{\sin\pr{\frac{n\pi x}{L_x}}}{n^3}.\end{align}
This result is presented in Fig.~\ref{fig:buckled-profile-rectangle}.
 
\section{Antiferromagnetic ground state for lattices without triangular loops}\label{app:GS-2d}

The partition function of the Ising model with nearest-neighbour interactions has only been analytically solved for the case of  one-dimensional and two-dimensional lattices~\cite{onsager_statistical_49}---in absence of external field for the latter. Nevertheless, it is possible to derive an expression of the low-temperature limit of the  partition function  for systems with specific properties, as shown below.   

For a general $d$-dimensional lattice, let the general antiferromagnetic Ising Hamiltonian be
\begin{equation}
    \mathcal{H}(\bm{\sigma})=-h\sum_{i}\space\sigma_i+J\sum_{<i,j>}\sigma_i\sigma_j, \qquad h>0,J>0,
\end{equation}
where $\sigma_i$ stands for the two-state spin variable at site $i$ and sum over  $<i,j>$ means sum over all nearest-neighbour pairs. It is handy to introduce the parameters defining the basic topological properties of the lattice, which are the number of spins $N$ and the coordination number $c$, the latter being the number of nearest neighbours of any site within the lattice. The energy of the system can be fully characterised through the number of spins aligned with the external field $n_h$, and the number of nearest-neighbour pairs with anti-aligned spins $n_J$. Equivalently, we may use the fractions $x_h\equiv n_h/N$ and $x_J=n_J/N$, which are particularly adequate to analyse the large system size limit $N\to\infty$. Taking into account this parametrisation, we have
\begin{align}
    \mathcal{H}\left(x_h,x_J\right)&=-h[n_h-(N-n_h)]+\!J\!\left[\left(\dfrac{c}{2}N-n_J\right)-n_J\right]
    \nonumber \\
    &=-N\left[ h(2x_h-1)+J\left(2x_J-\dfrac{c}{2} \right)\right].
    \label{eq:AppendixB_IsingEnergy}
\end{align}
In this way, a macrostate of the spin system is defined by a point $\left(x_h,x_J\right)$.

The partition function of the model is
\begin{equation}
    Z=\sum_{\hat{\sigma}} e^{-\beta \mathcal{H} (\bm{\sigma})}=\sum_{\left(x_h,x_J\right)} \xi \left(x_h,x_J\right) e^{-\beta \mathcal{H} \left(x_h,x_J\right)},
\end{equation}
where we have introduced the multiplicity $\xi\left(x_h,x_J\right)$ of the macrostate---i.e. the number of microstates compatible with the values $(x_h,x_J)$. In the low-temperature limit, $T\to 0^+$ ($\beta \to +\infty$), the leading order of the partition function stems from the ground state energy,
\begin {equation}
-\frac{1}{N\beta}\ln Z  \simeq e_{\gs}, \quad \beta\to\infty,
\end{equation}
where 
\begin{align}\label{eq:egs-general}
    e_{\gs} & \equiv \frac{1}{N}\min_{\left( x_h, x_J \right)} \mathcal{H}\left( x_h, x_J \right) \nonumber \\
     &= \min_{\left( x_h, x_J \right)} \left\{-h\left(2x_h-1\right)-J\left(2 x_J-\dfrac{c}{2} \right)\right\}
\end{align}
is the minimum energy of the system per lattice site. Therefore, the energy of the ground state plays the role of the free energy of the system in the low-temperature limit.
Since Eq.~\eqref{eq:AppendixB_IsingEnergy} is a linear function of the two variables $\left( x_h,x_J \right)$,  the minimum energy must be found at some point belonging to the boundary of the physically available set $\left( x_h,x_J \right)$.

\begin{figure}
    \centering
    \includegraphics[width=3.25in]{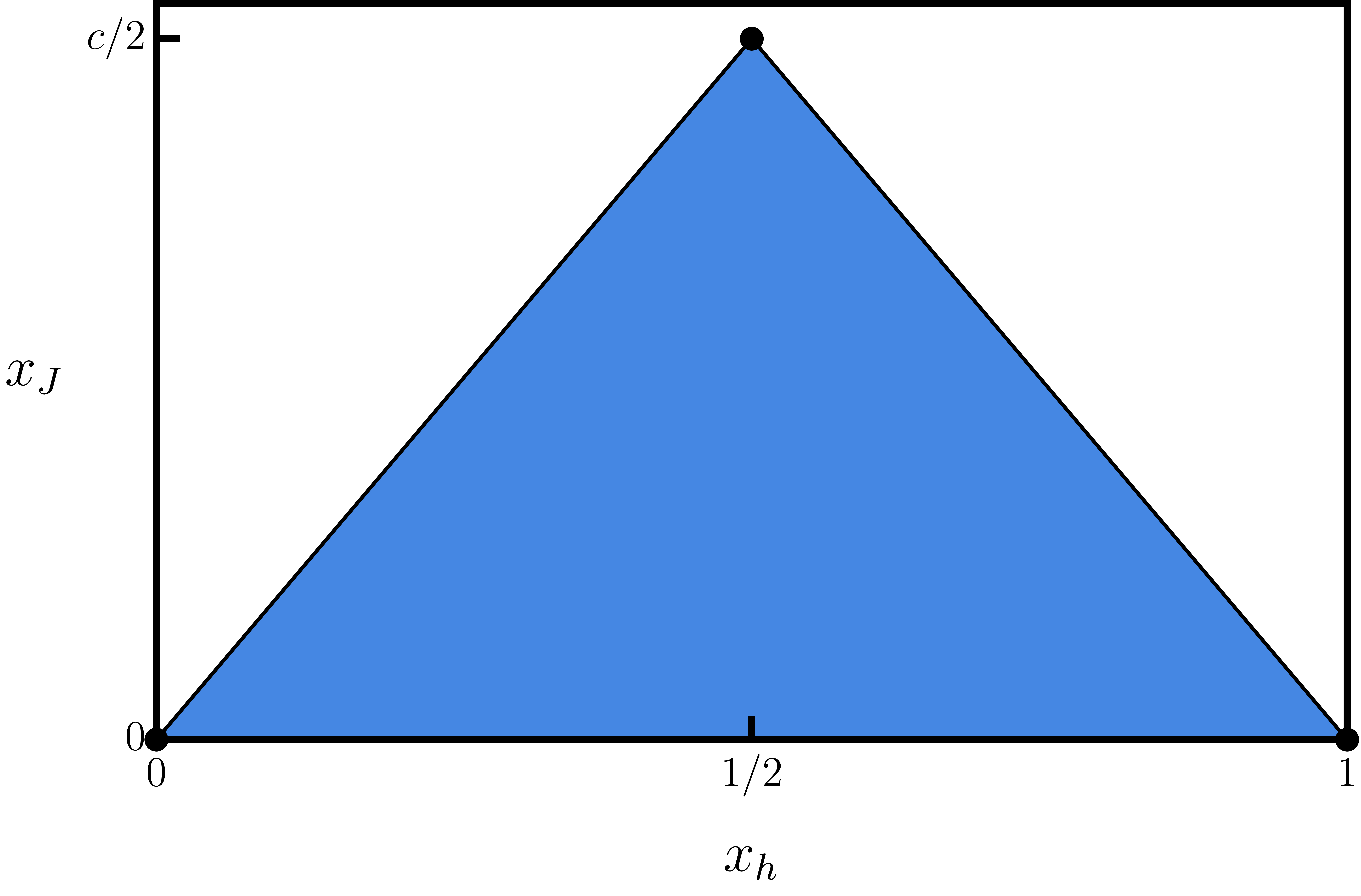}
    \caption{Antiferromagnetic Ising model parameter space $(x_h,x_J)$. The number of spins aligned with the external field is $n_h=N x_h$, whereas the number of antiferromagnetically aligned pairs in $n_J=N x_J$. The filled blue region indicates the physically available macrostates in the large system size limit as $N\to\infty$, for an arbitrary $d$-dimensional lattice without triangular loops.}
    \label{fig:AppendixB_ConfigurationalSpace}
\end{figure}

Note that $x_h$ is bounded both from below and from above by $0$ and $1$ respectively. Instead, $x_J$ is bounded  by $0$ and $c/2$. Nevertheless, not all points inside the rectangle $\left( x_h ,x_J \right) \in [0,1]\times [0,c/2]$ are physically acceptable. The specific shape of the physically available region in the $\left( x_h, x_J\right)$ space  is not straightforward to derive for arbitrary lattices. Nevertheless, let us restrict ourselves to the family of lattices with no triangular loops that contains, for instance, all bipartite networks---such as the honeycomb lattice in Fig.~\ref{fig:HoneycombLattice} or the rectangular lattice. The assumption of the absence of triangular loops allows to characterise exactly the available control set in the $\left( x_h, x_J\right)$ space. If this condition holds, the number of anti-aligned couples verifies 
\begin{equation}\label{eq:AppendixB_ConditionBipartite}
    x_J\leq c\ \text{min}(x_h,1-x_h).
\end{equation}
Hence, the three vertices $(0,0)$, $(1/2,c/2)$, and $(1,0)$ define a triangular region in the  $\left( x_h, x_J\right)$ plane in which all physically possible macrostates are contained. Moreover, in the large system size limit as $N\to \infty$, this region is densely filled by such macrostates---both $x_h$ and $x_J$ become continuous variables in this limit. 
The plane $(x_h,x_J)$ is shown in Fig.~\ref{fig:AppendixB_ConfigurationalSpace}, the blue region representing the physically available macrostates in the system---were we not working in the limit $N\to\infty$, only some points inside the blue region would represent acceptable macrostates.

After evaluating the energy at the boundaries, it is possible to obtain finally what is the energy of the ground state---and thus the low-temperature limit of the free energy of the system. Depending on the parameters $h$, $J$, there appear only two possibilities:
\begin{itemize}
    \item All spins are aligned with the external field for $h>cJ$. In this situation, the external field prevails over the ferromagnetic interaction. If so, $n_h=N$ and $n_J=0$, i.e. $x_h=1$ and $x_J=0$.
    \item The state is purely antiferromagnetic, i.e. all spins are antiparallel to their nearest neighbours, for $cJ>h$. In this regime, it is the antiferromagnetic coupling that dominates. Such a state exists due to our assuming that triangular loops are absent---otherwise, it would not be possible to have all nearest neighbours of any site $i$ antiparallel to $\sigma_i$, some of them would be nearest neighbours among themselves.  If so, $n_h=N/2$ and $n_J=cN/2$, i.e. $x_h=1/2$ and $x_J=c/2$.
\end{itemize}
Taking into account the above discussion, the energy per site in the ground state can be cast in a single equation,
\begin{equation}\label{eq:egs-expression-arbitrary-d} 
    e_{\gs} = -\left(h-c J\right) H(h-cJ)-\frac{c}{2}J,
\end{equation}
with $H$ denoting the Heaviside step function. This expression agrees with Eqs.~\eqref{eq:egs-expression-1d} and \eqref{eq:egs-expression-2d} in the main text for the one-dimensional and two-dimensional cases, where (i) the absolute value of the curvature $|\chi|$ plays the role of the external field,  and (ii) the coordination number is $c=2$ for $d=1$ and $c=3$ on the honeycomb lattice for $d=2$. Still, it must be remarked that Eq.~\eqref{eq:egs-expression-arbitrary-d} holds for an arbitrary $d$-dimensional lattice---as long as it does not contain triangular loops.

\section*{Acknowledgments}
We  acknowledge financial support from Grant No. PID2021-122588NB-I00 funded by MCIN/AEI/10.13039/501100011033 and by ``ERDF A way of making Europe.'' C.A.P. acknowledges the funding from European Union for the support from Horizon-Marie Sklodowska-Curie 2021 programme through the Postdoctoral Fellowship Ref.~101065902 (ORION). We also thank  Miguel Ruiz-García and Abigail Plummer for useful discussions.

\bibliography{Mi-biblioteca-14-dic-2022}

\end{document}